\documentclass[aps, prd, nofootinbib, superscriptaddress, twocolumn,]{revtex4-1}

\usepackage{amsmath,amssymb,amsthm,amstext,amsfonts}
\usepackage{graphicx}
\usepackage[english]{babel}
\usepackage{bm}
\usepackage[utf8]{inputenc}  
\usepackage{hyperref}

\def\begitem{\begin{itemize}}
\def\enditem{\end{itemize}}
\def\beq{\begin{equation}}
\def\eeq{\end{equation}}
\def\bey{\begin{eqnarray}}
\def\eey{\end{eqnarray}}

\newcommand{\non}{\nonumber}

\def\beq{\begin{equation}}
\def\eeq{\end{equation}}

\def\a0{$a_0$}

\def\phi{\varphi} 

\newcommand{\bg}{\bar\gamma}

\begin{document}

\title{Fab Four: When John and George play gravitation and cosmology}

\author{J.-P. Bruneton}
\email{jpbr@math.fundp.ac.be}
\affiliation{Namur Center for Complex systems (naXys),\\ University of Namur, Belgium}
\author{M. Rinaldi}
\email{mrinaldi@fundp.ac.be}
\affiliation{Namur Center for Complex systems (naXys),\\ University of Namur, Belgium}
\author{A. Kanfon}
\email{kanfon@yahoo.fr}
\affiliation{Facult\'e des Sciences et Techniques, \\
Universit\'e d'Abomey-Calavi\\
BP 526 Cotonou, B\'enin}
\author{A.~Hees}
\email{aurelien.hees@oma.be}
\affiliation{Royal Observatory of Belgium \\ Avenue Circulaire 3, 1180 Bruxelles, Belgium}
\affiliation{LNE-SYRTE, Observatoire de Paris, CNRS, UPMC\\ Avenue de l'Observatoire 61, 75014 Paris}
\affiliation{Namur Center for Complex systems (naXys),\\ University of Namur, Belgium}   
\author{S. Schl\"ogel}
\email{sandrine.schlogel@hotmail.com}
\affiliation{Namur Center for Complex systems (naXys),\\ University of Namur, Belgium}
\author{A.~F\"uzfa}
\email{andre.fuzfa@fundp.ac.be}
\affiliation{Namur Center for Complex systems (naXys),\\ University of Namur, Belgium}
\affiliation{Center for Cosmology, Particle Physics and Phenomenology (CP3), \\ University of Louvain, Belgium}

\begin{abstract}
\noindent Scalar-tensor theories of gravitation have recently regained a great interest after the discovery of the Chameleon mechanism and of the Galileon models. The former allows, in principle, to reconcile the presence of cosmological scalar fields with the constraints from experiments at the Solar System scale. The latter open up the possibility of building inflationary models that, among other things, do not need ad hoc potentials. Further generalizations have finally led to the most general tensor-scalar theory, recently dubbed the ``Fab Four'', with only first and second order derivatives of the fields in the equations of motion and that self-tune to a vanishing cosmological constant. This model has a very rich phenomenology that needs to be explored and confronted with experimental data in order to constrain a very large parameter space. In this paper, we present some results regarding a subset of the theory named ``John'', which corresponds to a non-minimal derivative coupling between the scalar field and the Einstein tensor in the action. We show that this coupling gives rise to an inflationary model with very unnatural initial conditions. Thus, we include a non-minimal, but non-derivative, coupling between scalar field and Ricci scalar, a term named ``George'' in the Fab Four terminology. In this way, we find a more sensible inflationary model, and, by performing a post-newtonian expansion of spherically symmetric solutions, we derive the set of equations that constrain the parameter space with data from experiments in the solar system.
\end{abstract}

\maketitle

\section{Introduction}

\noindent The Galileon theory has recently emerged as an effective theoretical realization of the Dvali-Gabadaze-Porrati model (DGP)  \cite{Nicolis:2008in}. The subsequent developments eventually led to the definition of the most general second order scalar-tensor theory that includes, besides the usual terms of scalar-tensor or $f(R)$ theories (as in, e.g. \cite{Damour:1992we, Sotiriou:2008rp}), also non-minimal derivative couplings to the curvature \cite{Deffayet:2009wt, Deffayet:2011gz}. This general theory, explored for the first time by the pioneering work of Horndeski many years ago \cite{Horndeski}, provides a wide framework that virtually encompasses all scalar models analyzed so far in the literature.

One interesting aspect of the non-linearities of the scalar sector is that they trigger the Vainshtein mechanism that makes possible to build viable cosmological models with sufficiently small effects at local scales to evade Solar System constraints.  Therefore, we have at hand an interesting alternative to chameleons \cite{Khoury:2003rn}, in which the parameter space for local gravity does not overlap with the one  allowed by  cosmic acceleration, as recently shown in \cite{Hees:2011mu}. In addition, inflationary phases are permitted without the introduction of ad hoc scalar potentials, making these models more ``natural''.
Non-linearities are also responsible for new phenomena in the dark sector including, for instance, sub/superluminality and/or effective violation of the Null Energy Condition, thus allowing for a stable and well-defined phantom-like phase \cite{Creminelli:2010ba, Babichev:2010kj}.

Generally speaking, the Galileon model has opened the way to new models for cosmology, including inflationary or late-accelerated ones \cite{Chow:2009fm, Charmousis:2011bf, Kobayashi:2011nu}. In fact, almost all sort of cosmological scenarios are possible depending on which Galileon model is chosen, see e.g. \cite{Saridakis:2010mf}.

In this paper, we wish to focus  on a subclass of models dubbed ``purely kinetic gravity" \cite{Sushkov:2009hk, Saridakis:2010mf, Gubitosi:2011sg, Germani:2010hd}, or also``John" in the ``Fab Four's'' terminology \cite{Charmousis:2011bf}. This also appears as a special case of the very general Galileon class considered in \cite{Tsujikawa:2012mk}.

We begin with a review of the equations of motion in a flat and empty Universe. We then complete the analysis made in \cite{Sushkov:2009hk} by computing the number of e-folds  in Sec. \ref{sec2b}. We show that a kinetically driven inflationary phase requires highly transplanckian values for the initial field velocity, which rule out the model. Moreover, in Sec.~\ref{sec2c} we provide for a detailed analysis of the no-ghost and causality conditions during cosmic evolution, and we show that the theory becomes acausal for such transplanckian values unless the coupling constant is vanishingly small or if the initial Hubble constant is transplanckian. In passing, we prove that some claims made in the literature regarding the sign of the coupling constant are wrong. We argue that this model is to be discarded also because there are no reasons why the scalar field should be generated at all in the first place. Indeed, although $\phi=0$ is only one possible solution of the equations of motion in the vacuum,   the theory is forced to reduce to pure GR when the same equations are solved inside a compact body.

In Sec.~III, we  extend our considerations to a more general model, in which we include a coupling of the scalar field to the Ricci scalar (and thus, eventually, to matter). This corresponds to the ``John plus George'' combination in the Fab Four terminology. The equations of motion, together with no-ghosts and causality conditions, are derived and studied numerically for a cosmological background. We find that inflation is possible provided both  coupling constants are positive. In Sec.~\ref{sec3c} we derive the field equations in spherical and static symmetry, and we put some constraints on the parameter space using Solar System tests. We conclude in Section IV with some remarks and open problems.

\section{John Lagrangian}
\label{sec2}
 \noindent We begin our analysis by considering the action 
\begin{eqnarray}
\label{actionjohn}
S&=&\int \sqrt{-g} d^4 x \left[\frac{R}{2 \kappa} - \frac{1}{2}\left(g^{\mu\nu}+ \kappa \gamma G^{\mu\nu}\right) \partial_{\mu}\phi\partial_{\nu}\phi\right]+\nonumber \\
&+& S_{\textrm{mat}}[\psi, g_{\mu\nu}],
\end{eqnarray}
where $R$ is the curvature scalar, $G^{\mu\nu}$ the Einstein tensor, $\psi$ collectively denotes the matter degrees of freedom coupled to the metric $g_{\mu\nu}$, and $\kappa=8 \pi G_{\rm N}$.  In this paper we use the mostly plus signature;  $\gamma$ is a dimensionless parameter whereas $\phi$ has the dimension of an inverse length. This action is a special case of the generalized Galileon one presented in \cite{Kobayashi:2011nu}, as can one can see by setting $K(X)=X, G_3=0, G_4=1/(2\kappa), G_5 = \kappa \gamma \phi/2$.

\subsection{Inflation with John}
\label{sec2b}
\begin{figure}[ht]
\begin{center}
\includegraphics[width=8cm]{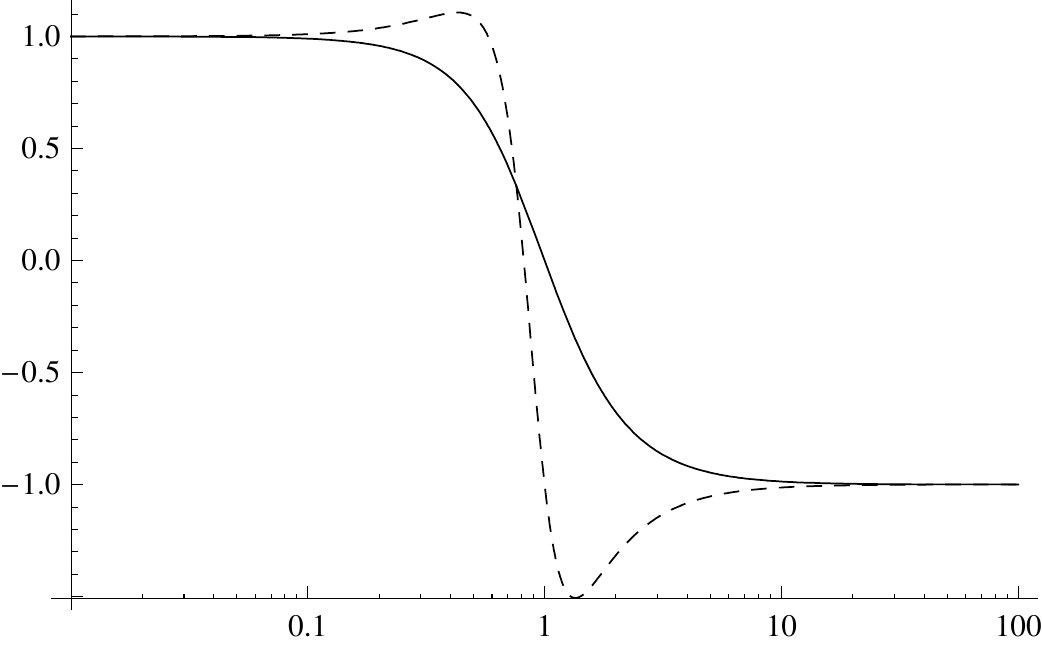}
\end{center}
\caption{Loglinear plot of $\omega_{\phi}$ as a function the dimensionless variable $\kappa \dot{\phi}$ for $\gamma=1$ and $\gamma=-1$ (dashed).}
\label{figwphi}
\end{figure}

\noindent As it was realized in \cite{Sushkov:2009hk}, this model allows for a quasi de Sitter  inflation with a graceful exit without the need for any specific scalar potential. Inflation is essentially driven by kinematics and it crucially depends on the initial high velocity of the field, as we will shortly see. Although, in principle, the  inflationary solutions begin at $t=-\infty$ (see \cite{Sushkov:2009hk}), we will  consider the action as an effective  model only  valid from few Planck times after an unknown transplanckian phase. Our first concern is to establish  whether the model accommodates an inflationary phase together with reasonable assumptions for the initial conditions at that time. This section thus complete the analysis found in \cite{Sushkov:2009hk} by providing the number of e-foldings as a function of the free parameters of the theory. The cosmological equations in vacuum derived from Eq.~(\ref{actionjohn}) read
\begin{subequations}\label{eom}
\begin{eqnarray}
&&3 \dot{\alpha}^2 = \frac{ \kappa\dot{\phi}^2}{2} \left( 1- 9 \kappa \gamma \dot{\alpha}^2\right), \label{eom1} \\
&&2\ddot{\alpha}+3\dot{\alpha}^{2}=-\frac{\kappa\dot{\phi}^{2}}{2}\left(1+\kappa \gamma\left[3\dot{\alpha}^{2}+2\ddot{\alpha}+4\dot{\alpha}\ddot{\phi}\dot{\phi}^{-1}\right]\right), \nonumber\\
\\
&&\frac{1}{a^3} \frac{d}{dt} \left(a^3 \dot{\phi}\left(1-3 \kappa \gamma \dot{\alpha}^2\right) \right)=0,
\end{eqnarray}
\end{subequations}
where $a$ is the scale factor and $\alpha = \ln a$. The system can be partially decoupled to allow for a numerical integration whose results are shown in Sec.~\ref{sec2numres}. Isolating the second order derivatives, we find
\begin{subequations}\label{eomdecoupled}
\begin{eqnarray}
\ddot{\alpha}&=&\frac{\left( 3 \kappa \gamma \dot{\alpha}^2 -1\right)}{2}\frac{3 \dot{\alpha}^2 +  \frac{\kappa \dot{\phi}^2}{2}\left(1-9\kappa \gamma \dot{\alpha}^2 \right) } {1-3\gamma \kappa \dot{\alpha}^2+ \frac{\kappa^2 \gamma \dot{\phi}^2}{2} \left(1+9\kappa \gamma \dot{\alpha}^2 \right)}, \nonumber  \label{eomd1} \\
\\
\ddot{\phi}&=&\frac{-3 \dot{\alpha} \dot{\phi} \left(1+\kappa^2 \gamma \dot{\phi}^2\right)}{1-3\gamma \kappa \dot{\alpha}^2+ \frac{\kappa^2 \gamma \dot{\phi}^2}{2} \left(1+9\kappa \gamma \dot{\alpha}^2 \right)},\label{eomd2}
\end{eqnarray}
\end{subequations}
which can be integrated numerically in a straightforward way. The effective equation of state (EoS) for the scalar field can be obtained from  its stress-energy tensor or, more simply, by comparing our equations of motion directly to the standard Friedmann equations. After some algebra we find
\beq
\label{wphi}
\omega_{\phi}=\frac{\left(2+3\kappa^2 \gamma\dot{\phi}^{2}\right)\left(1-\kappa^2\gamma\dot{\phi}^{2}\right)}{2+3\kappa^2\gamma\dot{\phi}^{2}+3\kappa^{4}\gamma^{2}\dot{\phi}^{4}},
\eeq
which is plotted in Fig.~\ref{figwphi} for both positive and negative $\gamma$. For both signs of $\gamma$, the EoS tends to $-1$ in the high energy limit ($\kappa |\dot{\phi}| \gg 1$), so that a large initial velocity for the scalar field will result in a quasi de Sitter phase.  However, only the case of positive $\gamma$ can lead to inflation. Indeed,  Eq.~(\ref{eom1}) can be inverted to $3 \dot{\alpha}^2=\kappa \dot{\phi}^2/(2+3 \kappa ^2\gamma \dot{\phi}^2)$, which needs to be positive since the Hubble constant is a real number. Thus, $\gamma <0$ implies that $|\dot{\phi}|< \sqrt{-2/3\gamma}$, which, in turn, means that we always have $w_{\phi}>0$. Therefore, the scalar field cannot even start in the $w_{\phi} <0 $ region if $\gamma <0$. More generally speaking, accelerated phases driven by a scalar field in this model require $\gamma>0$.

In view of these considerations, in the following we shall assume that $\gamma$ is positive.  Eq.~(\ref{eomd2}), together with the condition that $1-3 \kappa \gamma \dot{\alpha}^2 >0$ (see Sec.~\ref{sec2c}), shows that $\ddot{\phi} < 0$ for an initially expanding Universe ($\dot{\alpha}>0$). Hence, the velocity of the field decreases with time and $\omega_{\phi}$ is driven towards $\omega =+1$. We characterize the end of inflation by the instant $t_{\text{end}}$, at which  $\omega_{\phi}=-1/3$ \footnote{Throughout this paper we assume a vanishing cosmological constant.}.
Under the assumption that $\kappa \gamma \dot{\phi}^2$ is initially large, one may derive an analytical (approximate) solution for the scale factor $a(t)=e^{\alpha(t)}$ and the scalar field at early times. First, we write Eq.~(\ref{eom1}) as $3 \dot{\alpha}^2= \kappa \dot{\phi}^2/(2 + 3 \kappa^2 \gamma \dot{\phi}^2)$ so that 
\beq
\label{Hhigh}
H=\dot{\alpha} \simeq \frac{1}{3 \sqrt{\kappa \gamma}},
\eeq
where $H$ is the Hubble constant. Integration yields the approximate scale factor
\beq
\label{scalefactor}
a \sim  a_i \exp\left( \frac{t -t_i}{3 \sqrt{\kappa \gamma}}\right).
\eeq
Using Eq.~(\ref{Hhigh}) in Eq.~(\ref{eomd2}), and expanding according to $\kappa^2 \dot{\phi}^2 \gg 1$, gives $\ddot{\phi}/\dot{\phi} \simeq -1/\sqrt{\kappa \gamma}$ and 
\beq
\label{phidot}
\dot{\phi} \sim \dot{\phi}_i \exp\left(-\frac{t-t_i}{\sqrt{\kappa \gamma}}\right).
\eeq
Now recall that  inflation ends at $t_{\text{end}}$ such that $\omega_{\phi}(t_{\text{end}}) =-1/3$. This corresponds to $\kappa^2 \gamma \dot{\phi}^2 = \zeta_{\text{end}}$, where 
\beq
\zeta_{end} = \frac{1}{6} \left(3+\sqrt{57}\right) \approx 1.76,
\eeq
as can be shown by solving Eq.~(\ref{wphi}). Using Eq.~(\ref{phidot}), the condition $\kappa^2 \gamma \dot{\phi}^2 = \zeta_{\text{end}}$ reduces to
\beq
\kappa^2\gamma\dot{\phi}_{i}^{2}\exp\left(-\frac{2 (t_{end} - t_i)}{\sqrt{\kappa \gamma}}\right) \sim \zeta_{\text{end}},
\eeq
from which one finds
\beq
\label{deltat}
t_{end}-t_i=\frac{\sqrt{\kappa \gamma}}{2}\ln \left(\frac{\kappa^2\gamma\dot{\phi}_{i}^{2}}{\zeta_{\text{end}}}\right).
\eeq
Replacing this in the expression (\ref{scalefactor}) for the scale factor leads to
\beq
\label{fraca}
\frac{a_{\text{end}}}{a_i} \sim \left(\frac{\kappa^2 \gamma \dot{\phi}_i^2}{\zeta_{\text{end}}}\right)^{\frac{1}{6}}.
\eeq
We finally impose that inflation lasts for a  number of e-folds $N= \ln (a_{\text{end}}/a_i)$ greater than $60$. This gives a relation between the initial velocity of the field $\dot{\phi}_i$ and $\gamma$, namely
\beq
\label{maineq}
\dot{\phi_i}^2 \gtrsim \frac{\zeta_{\text{end}}}{\kappa^2 \gamma} \exp(360),
\eeq
which is the crucial condition for a successful (purely kinetic-driven) inflationary phase. We see that it involves a rather unusual very large pure number. In order to discuss naturalness, Eq.~(\ref{Hhigh}) is also of interest, as it fixes the Hubble constant at the beginning of the inflationary phase $H_{i} \sim 1/ 3\sqrt{\kappa \gamma}$. Therefore, the last equation might also be written as
\beq
\kappa \frac {\dot{\phi_i}^2}{H_i^2} \gtrsim 9 \, \zeta_{\text{end}}\exp(360) \sim 10^{157}.
\eeq
It follows that the ``natural'' initial conditions $H_{i} = \mathcal{O}(1)$ and $\dot{\phi}_i  = \mathcal{O}(1)$ in Planckian units are not allowed. On the contrary, a natural value for the initial expansion  $H_{i} =1$ (and thus  $\gamma \approx 0.11$) requires an extremely high transplanckian value for the initial velocity of the field $\dot{\phi}_i \sim 10^{78}$ in natural units. 

It is not even possible to obtain a Planckian value for the initial velocity in this model, since, in any event, the initial Hubble constant will be greater than the one today. This implies that in such an inflationary scenario, $\sqrt{\kappa \gamma}$ must be less than the Hubble radius today, still implying a very unnatural  bound for the initial velocity, namely $\dot{\phi}_i \gtrsim 10^{51}$ in natural units. 

\subsection{Theoretical constraints}
\label{sec2c}
\noindent The characteristic feature of Galileon models is the derivative coupling of the scalar field to the metric. This implies a direct coupling between scalar field and metric degrees of freedom, or, in other words, that the scalar field propagation explicitly depends on the metric background and vice-versa. Therefore, there might exist backgrounds for which the propagation becomes pathological (non hyperbolic, ie. non causal, or carrying negative energy, ie. ghosts). In the following, we restrict ourselves to the cosmological background in a flat universe with line element
\beq
\label{lineelement}
ds^2= -dt^2 + e^{2 \alpha(t)} d\mathbf{x}^2,
\eeq 
and we explore the conditions for the theory to be well-defined, for both scalar field and metric perturbations. We start with the scalar field, whose action can  be written as 
\begin{equation}
S_{\phi}= \int a^3 dt d^3x Q_{\phi}\left( \dot{\phi}^2 - \frac{c_{\phi}^2}{a^2} \nabla \phi^2\right),
\end{equation}
where
\begin{equation}
\label{qphi}
Q_{\phi}=\frac{1}{2}\left(1-3 \kappa \gamma \dot{\alpha}^2 \right) >0
\end{equation}
as it needs to be positive for the scalar field to carry positive energy. The field propagates at a squared speed given by
\begin{equation}
\label{cphi}
c_{\phi}^2= \frac{1 - \kappa \gamma \left(2 \ddot{\alpha} + 3 \dot{\alpha}^2 \right)}{1-3 \kappa \gamma \dot{\alpha}^2 } \geq 0.
\end{equation}
The condition $c_{\phi}^2 \geq 0$ is necessary to ensure that the scalar field equation of motion remains hyperbolic,  ultimately expressing its causal behavior regardless of whether the scalar field perturbations are sub or superluminal \cite{Bruneton:2006gf,Babichev:2007dw}.  The two conditions above are indeed equivalent to the requirement that the effective metric \begin{eqnarray}
&&\tilde{g}^{\mu\nu}=g^{\mu\nu}+ \kappa\gamma G^{\mu\nu} \\
&&\tilde{g}^{\mu\nu} \nabla_{\mu \nu} \phi = 0,
\end{eqnarray}
along which the scalar field propagates, is hyperbolic with the same (mostly $+$) signature than $g_{\mu\nu}$. 

The conditions (\ref{qphi}) and (\ref{cphi}) are best analyzed in terms of the reduced dimensionless variables $x(t)= \kappa \dot{\phi}$ and $y(t)=\sqrt{\kappa} \dot{\alpha}$. The first one  requires $\gamma y^2<1/3$ and it is  automatically satisfied if $\gamma <0$, but also if $\gamma >0$ provided that there is a maximal value for the Hubble constant $H=\dot{\alpha} < 1/\sqrt{3 \kappa\gamma}$. The second one reduces to an algebraic condition with the help of  Eqs.~(\ref{eom}) and (\ref{eomdecoupled}), namely
\beq
\frac{1-3 \gamma y^2}{1-9 \gamma y^2+54 \gamma^2 y^4}\geq0 \Leftrightarrow \frac{\left(1+\gamma x^2\right) \left(2+3 \gamma x^2\right)}{2+3 \gamma x^2+3 \gamma^2 x^4}\geq0.
\eeq
In summary, Eqs.~(\ref{qphi}--\ref{cphi}) require $\gamma >0$ and $|y|<1/\sqrt{3 \gamma}$ or $\gamma<0$ and the two possibilities $|x|>1/\sqrt{-\gamma}$ or $|x|< \sqrt{-2/3\gamma}$. These conditions are therefore less restrictive than the one implied by the Friedmann equation (see previous section).

The curvature background implies non-standard propagation for the scalar degree of freedom. In a quite similar fashion, the scalar field background modifies the standard spectrum of metric perturbations. Similar conditions for the avoidance of ghosts and euclidean metrics also exist. These have been derived in full generality in a very wide class of Galileon models in \cite{DeFelice:2011bh}, whose conventions  we follow. These conditions, namely Eqs.~(23), (25--27) of \cite{DeFelice:2011bh} reduce to rather simple algebraic constraints in our case, after the necessary manipulation using the equations of motion (\ref{eom}) and (\ref{eomdecoupled})
\begin{subequations}
\begin{eqnarray}
Q_T>0   \Leftrightarrow 1+\frac{\gamma x^2}{2} >0, \\
c_T^2 \geq 0 \Leftrightarrow 1-\frac{\gamma x^2}{2}  \geq 0,
\end{eqnarray}
\end{subequations}
where $Q_T$ and $c_T$ are defined as their scalar counterpart $Q_{\phi}$ and $c_{\phi}^2$ above, but stand for the tensor perturbations of the metric field. Similar conditions need to hold for the scalar part of the metric perturbations, namely
\begin{subequations}
\begin{eqnarray}
&&Q_S>0   \Leftrightarrow \frac{4+6 \gamma x^2+6 \gamma^2 x^4}{2+3 \gamma x^2} >0,\\
&&c_S^2 \geq 0 \Leftrightarrow \frac{12+36 \gamma x^2+19 \gamma^2 x^4-12 \gamma^3 x^6-3 \gamma^4 x^8}{2+3 \gamma x^2+3 \gamma^2 x^4} \geq 0. \nonumber \\
\end{eqnarray}
\end{subequations}
This whole set of equations is difficult to reduce algebraically because of the last one. However, one might easily plot the six functions of $x$ defined above, and one typically finds that both  positive and negative values for $\gamma$ are allowed on a given range $|x|< \xi_{\gamma}$, where typically $\xi_{\gamma}$ behaves as $\mathcal{O}\left(1/\sqrt{|\gamma|}\right)$, see e.g. Figs. \ref{ghost1} and Fig. \ref{ghost2} below. Hence,  large (transplanckian) values for $|x|$ are only allowed for small $|\gamma| \ll 1$. This means that the space for possible velocities of the field $x=\kappa \dot{\phi}$ needs to be typically subplanckian, unless $\gamma$ is vanishingly small. This will be linked to the results found earlier, where transplanckian initial velocity were required for a successful inflation,  leading to negative squared speeds $c_{S}^2$ and $c_T^2$ in that epoch. This is shown in Fig.~\ref{johnnum2}.
\begin{figure}[ht]
\begin{center}
\includegraphics[width=7cm]{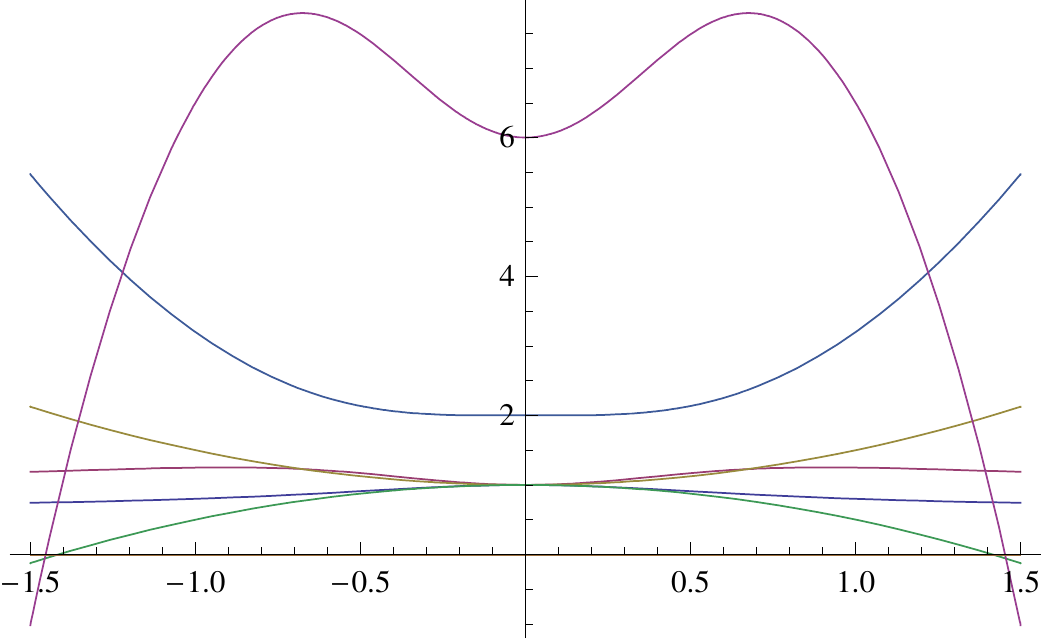}
\end{center}
\caption{Plot of the six conditions $Q_i$ and $c_i$ derived above, as a function $x$. Here $\gamma =1$. Allowed values for the field velocity is typically $|x|<\mathcal{O}(\gamma^{-1/2})$, as made clear by drawing similar plots while varying $\gamma$.}
\label{ghost1}
\end{figure}
\begin{figure}[ht]
\begin{center}
\includegraphics[width=7cm]{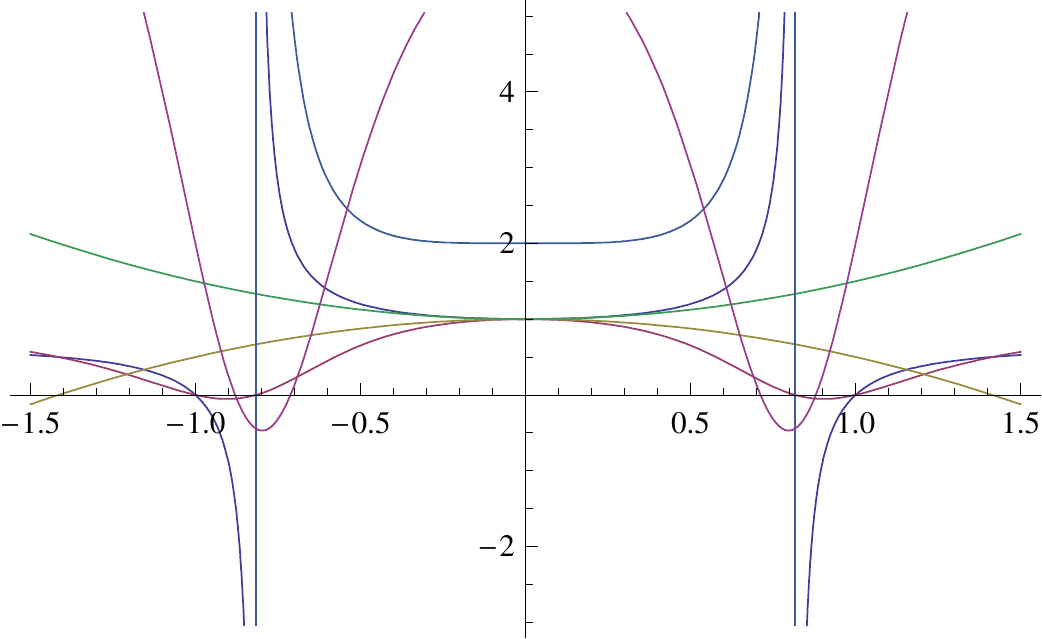}
\end{center}
\caption{Plot of the six conditions $Q_i$ and $c_i$ derived above, as a function $x$. Here $\gamma =-1$. Allowed values for the field velocity is typically $|x|<\mathcal{O}(|\gamma|^{-1/2})$.}
\label{ghost2}
\end{figure}

In passing, we note that the claim made in the literature (see e.g. \cite{Tsujikawa:2012mk, Germani:2010gm, Germani:2010hd}) according to which only the subclass $\gamma<0$ is a ghost-free theory is wrong (at least in the background considered here).  Notice that the scalar field is well-defined although being a phantom in certain regime (in the case $\gamma<0$), a situation reminiscent of the one discussed in \cite{Creminelli:2010ba}. However, as shown previously, the Friedmann equation actually prevents the scalar field to enter this regime.
%

\subsection{Numerical results}
\label{sec2numres}
\begin{figure}[ht]
\begin{center}
\includegraphics[width=8cm]{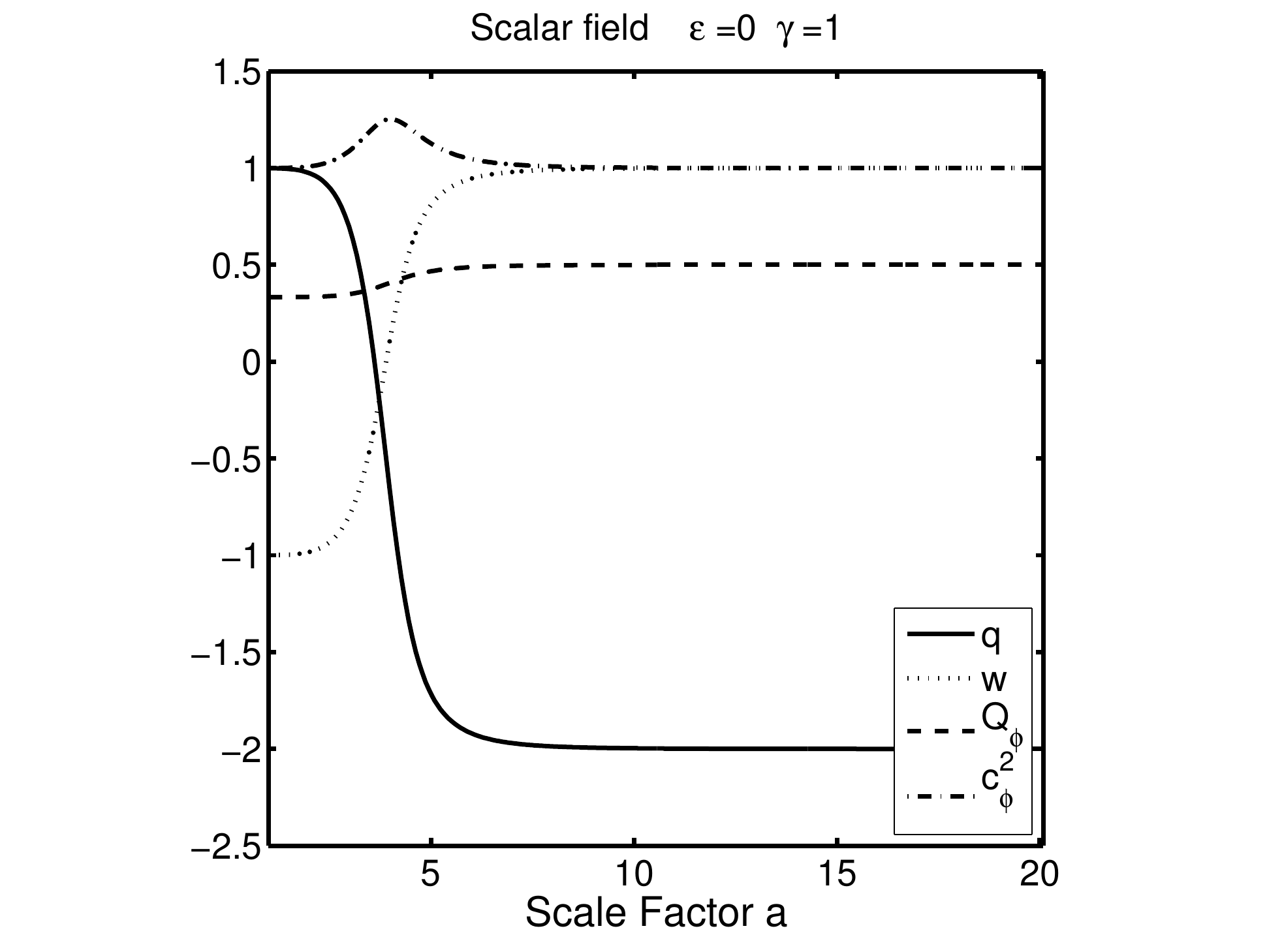}
\end{center}
\caption{Evolution of the acceleration parameter $q=+ \ddot{a} a/ \dot{a}^2$ vs. the scale factor, for $\gamma=1$. Also shown in the graph:  the evolution of the scalar field EoS $w_{\phi}$ and $Q_{\phi}$ and $c_{\phi}^2$. The scalar field becomes superluminal during the transition from a de Sitter Universe to a to stiff matter-dominated one. However, as discussed in the text, it remains hyperbolic and, thus, causal. Moreover, it carries positive energy.}
\label{johnnum1}
\end{figure}

\begin{figure}[ht]
\begin{center}
\includegraphics[width=8cm]{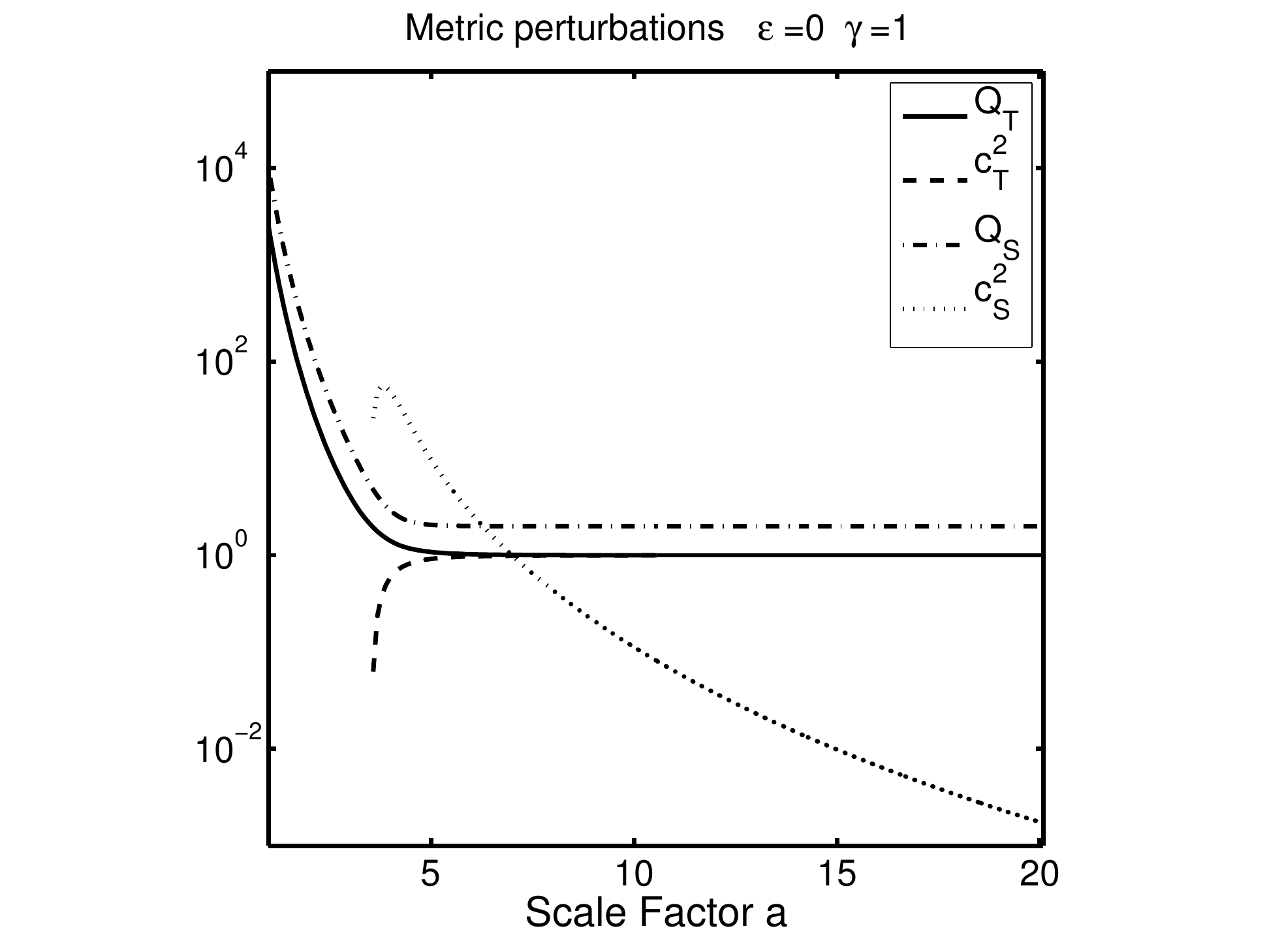}
\end{center}
\caption{Evolution of the parameters for metric perturbations (scalar and tensorial parts), respectively $Q_{S}$, $c_{S}^2$, $Q_{T}$, $c_{T}^2$, as defined in the text vs the scale factor, for $\gamma=1$. As found theoretically, the initial high velocity of the field drives both the speed of (scalar and tensor) metric perturbations to imaginary values, thus signaling a breakdown of hyperbolicity for metric perturbations. (Here it happens when the corresponding curves terminate, since the $y$ axis is in logarithmic scale).}
\label{johnnum2}
\end{figure}

\begin{figure}[ht]
\begin{center}
\includegraphics[width=8cm]{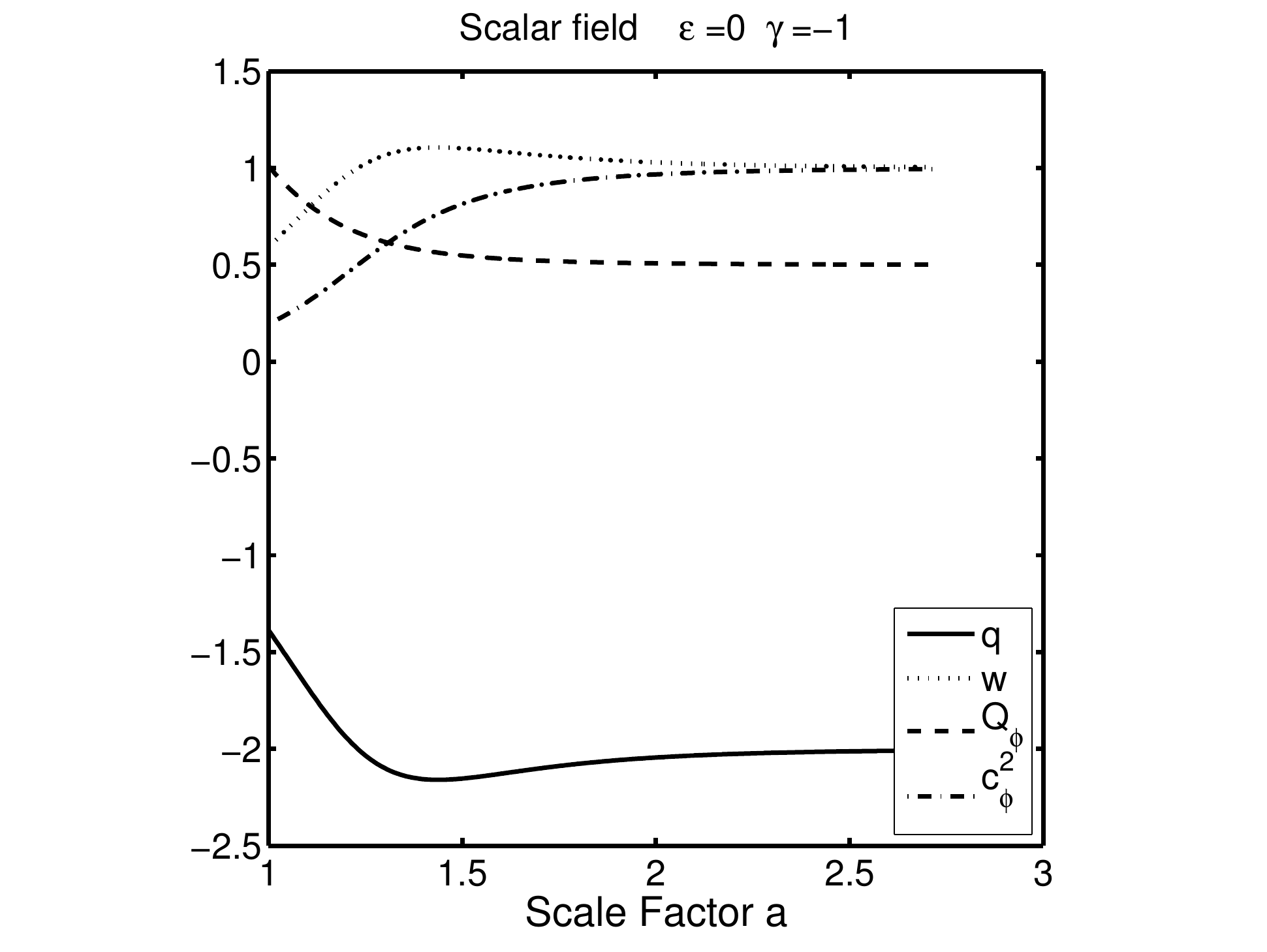}
\end{center}
\caption{Evolution of the acceleration parameter $q$ vs. the scale factor, for $\gamma=-1$ with initial condition satisfying  $|x_i|=0.1<\sqrt{-2/3\gamma}$. The field starts with an EoS $w_{\phi} \sim 0.5$ and the Universe only decelerates.}
\label{johnnum3}
\end{figure}

\begin{figure}[ht]
\begin{center}
\includegraphics[width=8cm]{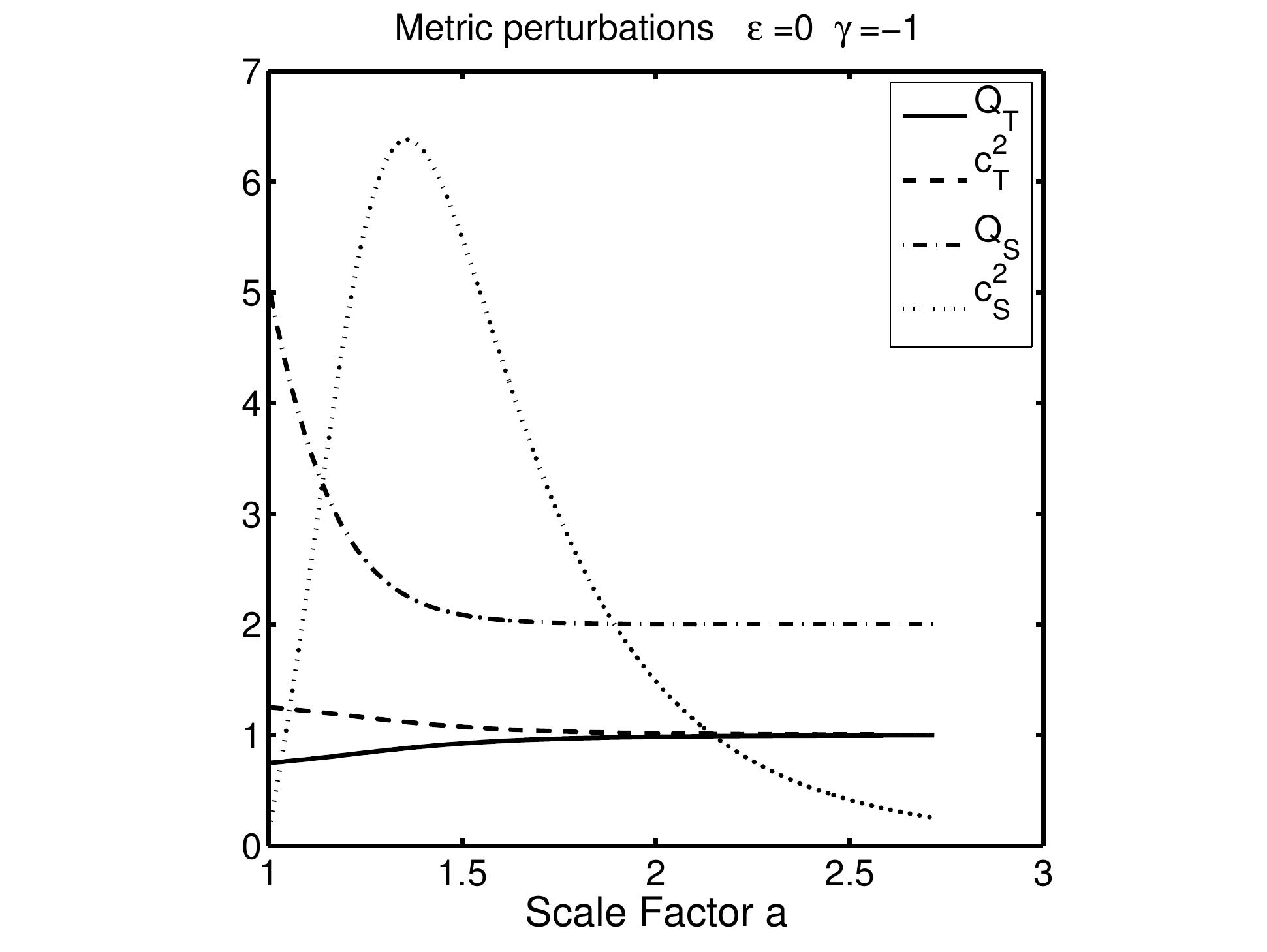}
\end{center}
\caption{Evolution of the parameters for metric perturbations (scalar and tensorial parts), respectively $Q_{S}$, $c_{S}^2$, $Q_{T}$, $c_{T}^2$, as defined in the text vs the scale factor, for $\gamma=-1$. This model with $\gamma<0$ is well-behaved but do not accommodate inflation.}
\label{johnnum4}
\end{figure}

\noindent In this section, we quickly show the cosmological behavior in the John model, for both positive and negative $\gamma$. As discussed before, the negative $\gamma$ case leads only to a decelerating Universe: the phantom regime is not an acceptable initial condition (as it entails an imaginary Hubble constant), and neither can be reached. Only  positive $\gamma$ leads to acceleration, and to an inflationary phase in the early Universe, a drawback being the presence of non-causal behavior for the scalar and tensor perturbations of the metric. These plots (Figs.~\ref{johnnum1}--\ref{johnnum4}) have been obtained by numerical integration for an initial condition of $\dot{\phi}_i =10$ in natural units in the case $\gamma=1$, and $\dot{\phi}_i =0.1$ for $\gamma=-1$. 

\subsection{Discussion}
\label{sec2d}
We have established that kinetically driven inflation in the Galileon theory involving the simplest coupling to Einstein tensor is not viable. It requires unnatural transplanckian values for the initial velocity of the field, which, in turn, implies various instabilities, as discussed above. 

This model has anyway another serious drawback. In the absence of any direct coupling to the Ricci scalar, there is no reason why the scalar field  should be generated at all (even in presence of cosmological matter fluid). In other words, $\phi =0$ is always a solution in this class of models, whatever the matter content is. At local scales, this problem appears in the following way. We checked numerically that a relativistic star (with flat asymptotic conditions) must be described by pure GR, i.e.  $\phi=0$ for all $r$ is the only solution that is regular at the center of the star. To conclude, the model considered so far is trivial in the sense that it cannot be different than GR, except if one imposes non-vanishing initial conditions for the scalar field at early times.

This acts as a leitmotiv for a more realistic model, studied in the next section, where we add to the Lagrangian a direct coupling to the Ricci scalar: $R(1+\epsilon \phi)$, namely the "George" term in the Fab Four terminology. 
\section{George and John}
\label{sec3}
We now consider the extended model given by  
\begin{eqnarray}
\label{actionjohngeorges}
S&=&\int \sqrt{-g} d^4 x \left[\frac{R}{2 \kappa}\left(1+\epsilon \sqrt{\kappa}\phi\right)+ \right. \\\non 
&& \left. - \frac{1}{2}\left(g^{\mu\nu}+ \kappa \gamma G^{\mu\nu}\right) \partial_{\mu}\phi\partial_{\nu}\phi\right]+ S_{\textrm{mat}}[\psi, g_{\mu\nu}] ,
\end{eqnarray}
where $\epsilon$ is a dimensionless, free parameter. Of course this is not the most general coupling one might consider but it is anyway reminiscent of Brans-Dicke coupling. Notice that one might worry that the effective gravitational constant $G_{\textrm{eff}} = G/(1+\epsilon \sqrt{\kappa} \phi)$ might easily become negative in this model, meaning that the action chosen here shall trivially lead to dynamical pathologies for $\epsilon \phi$ sufficiently large and negative\footnote{In fact, what matters in the case $\gamma=0$, is that the scalar field propagates positive energy in the Einstein frame. Performing a conformal transformation, this is equivalent to the usual Brans-Dicke condition $2 w + 3 >0$, where $w= \epsilon^2(1+ \epsilon \sqrt{\kappa} \phi)$ here. Then, our model with $\gamma =0$ would indeed be pathological if $\phi \leq -(3/(2\epsilon^2)+1)/(\epsilon \sqrt{\kappa})$ . However the $\gamma$ term introduces new terms in the equation of motion for the scalar field which invalidate such a conclusion in the general case $\gamma \neq 0$.}. 

Such an argument would call in favor of defining a better coupling function $F$ in the Georges term $F(\phi) R$. However, this would be a misleading conclusion here, since the John term couples the derivatives of the metric and of the scalar field, thus impacting their  propagation. Therefore, only the entire set of no-ghost conditions together with causal propagation conditions (positivity of the squared velocities) for both the scalar and the metric perturbations can decide which regions of the configuration space are well-behaved. This is done in the following sections (on a cosmological background), based on the conditions derived in Appendix \ref{AppB}. In this light, the function $F$ chosen above is just the simplest one could have chosen, and might furthermore be understood as retaining only the first term in a series expansion of a more general function $F$.

The cosmological evolution in this theory is typically a function of four parameters, the initial value of the field, its velocity, and of the two dimensionless parameters $\gamma$ and $\epsilon$. It goes beyond the scope of the present paper to provide with a comprehensive study of this parameter space. However we highlight some essential features of the model thanks to numerical results displayed in the next section. We provide both the cosmological evolution and the analysis of causality and positivity of energy within subclasses of the model, depending on the signs of $\gamma$ and $\epsilon$.

\subsection{Cosmological behavior}
\label{sec3b}
The equations of motion in a flat, empty Universe, derived from Eq.~(\ref{actionjohngeorges}) are given in Appendix \ref{AppA}, see Eqs.~(\ref{Eomjg1}-\ref{Eomjg3}). We extended the analysis of the no-ghost and causality conditions to this more general framework, and we also provide the scalar field EoS, see Appendix \ref{AppA} and \ref{AppB}. 
                  
The numerical results are the following. The case $\epsilon=1$ and $\gamma=1$ is pretty similar to the case John alone, see Figs.~(\ref{jgs11}) and (\ref{jgm11}) below. 
\begin{figure}[hbt]
\begin{center}
\includegraphics[width=0.35\textwidth]{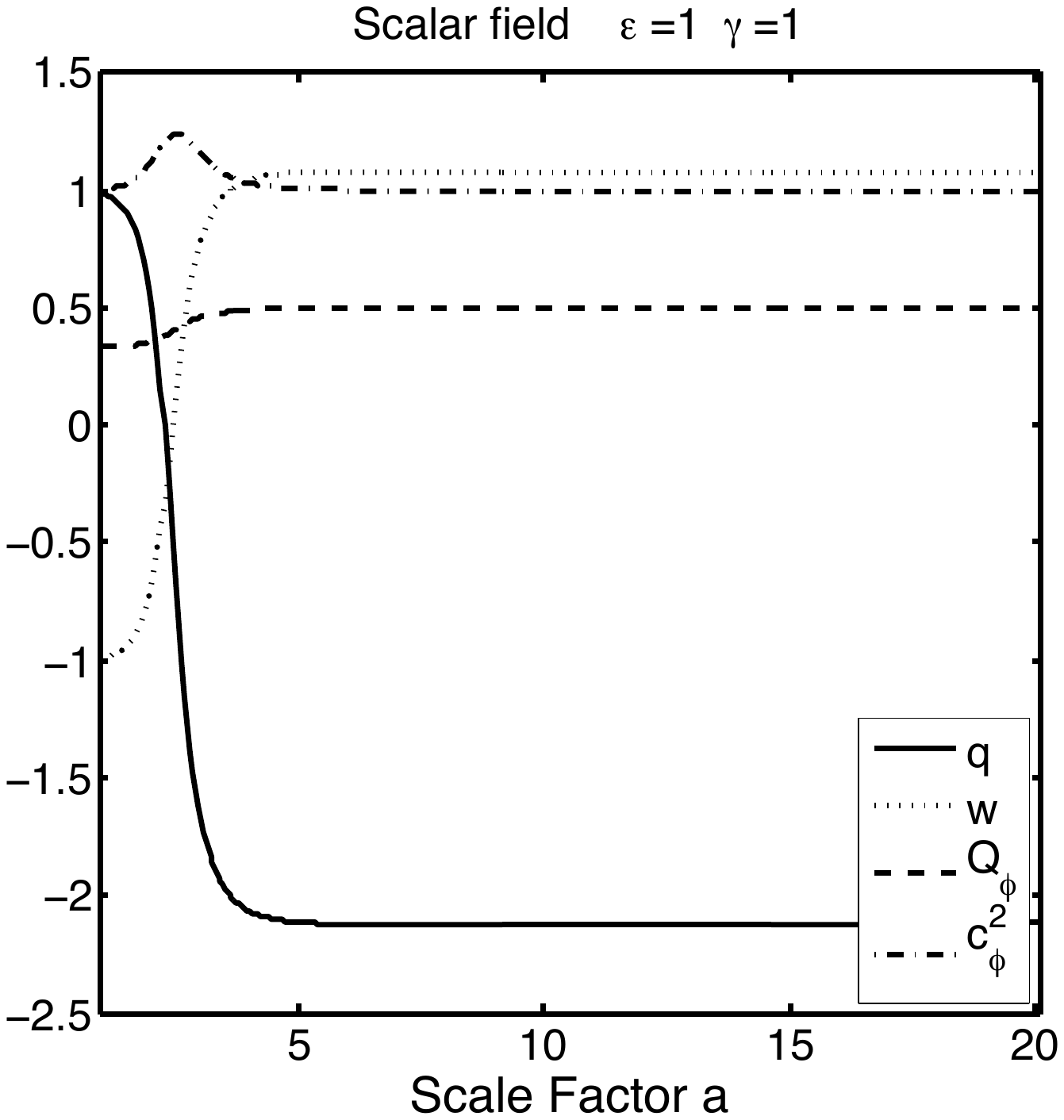}
\end{center}
\caption{Evolution of $q$, $w_{\phi}$, $Q_{\phi}$ and $c_{\phi}^2$ with the scale factor. We observe the same transition from inflation to stiff matter for the scalar field. Initial conditions are $\dot{\phi}_i=100$ and $\phi_{i}=1$.}
\label{jgs11}
\end{figure}
\begin{figure}[ht]
\begin{center}
\includegraphics[width=0.35\textwidth]{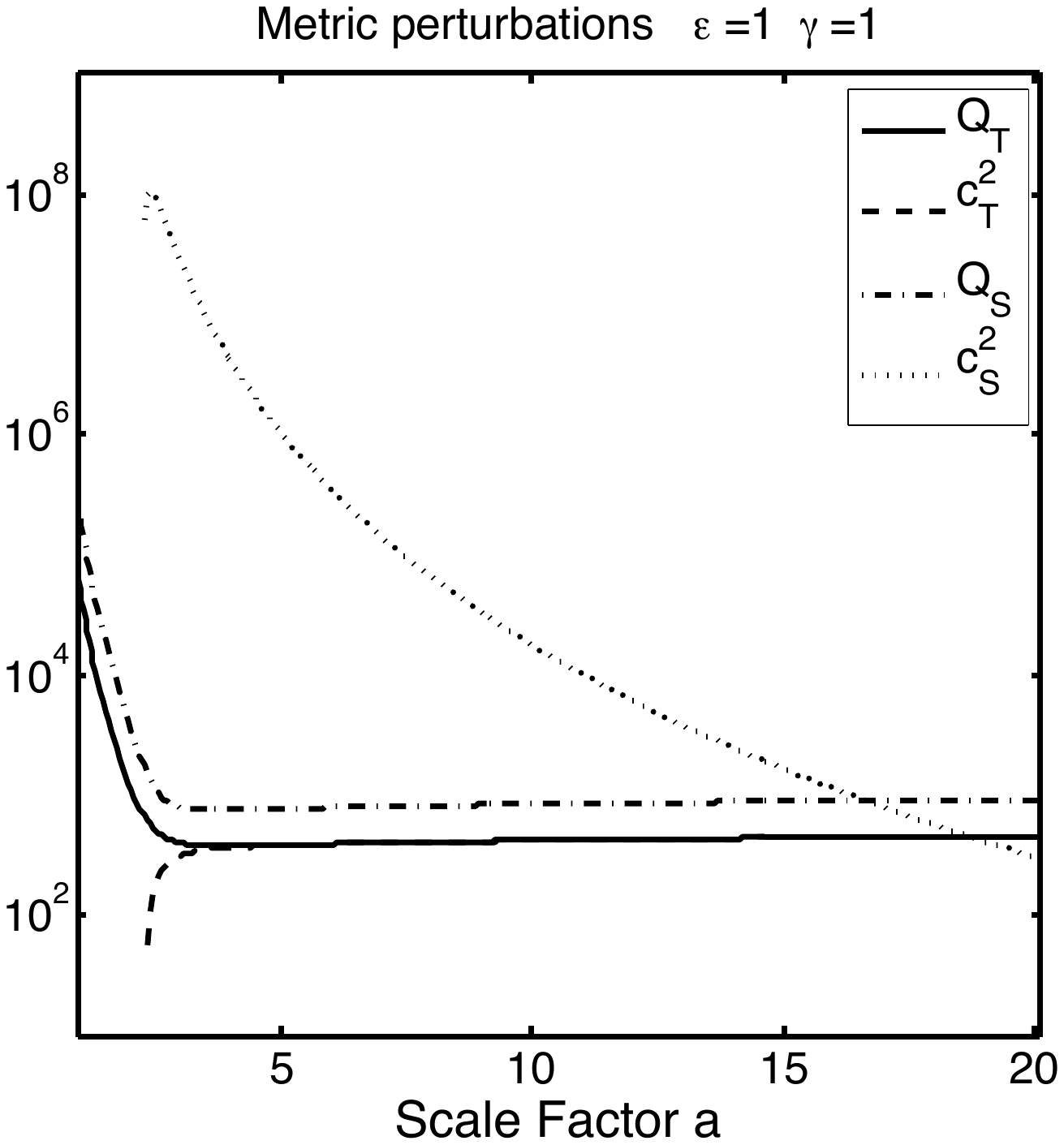}
\end{center}
\caption{Evolution of $Q_{T}$, $Q_{\phi}$ and $c_T^2, c_{\phi}^2$ with the scale factor. These two speeds are negative in the early universe.}
\label{jgm11}
\end{figure}   
Inflation thus occurs in the case $\epsilon>0$ and $\gamma>0$, but the acausal behavior still shows up in the very early Universe. The number of e-folds is a function of the two initial conditions for the field and its velocity, and the dimensionless parameters $\epsilon$ and $\gamma$. A further analysis that goes beyond te scope of this paper would determine whether the addition of the George term help in solving the naturalness problem encountered with John alone in Sec.~\ref{sec2}.

\begin{figure*}[bht]
\begin{center}
\includegraphics[width=0.35\textwidth]{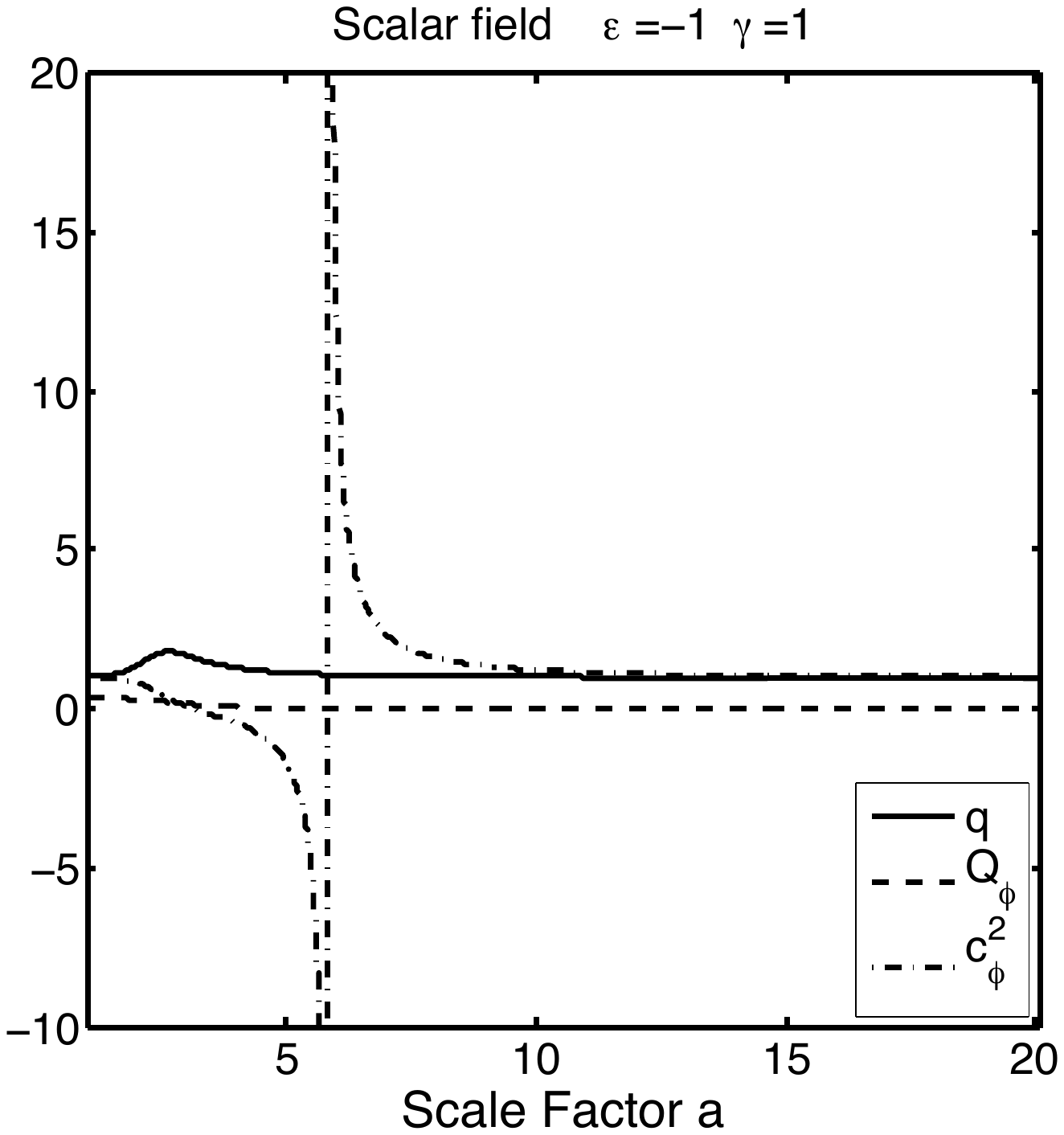}\hspace{1.5cm}
\includegraphics[width=0.35\textwidth]{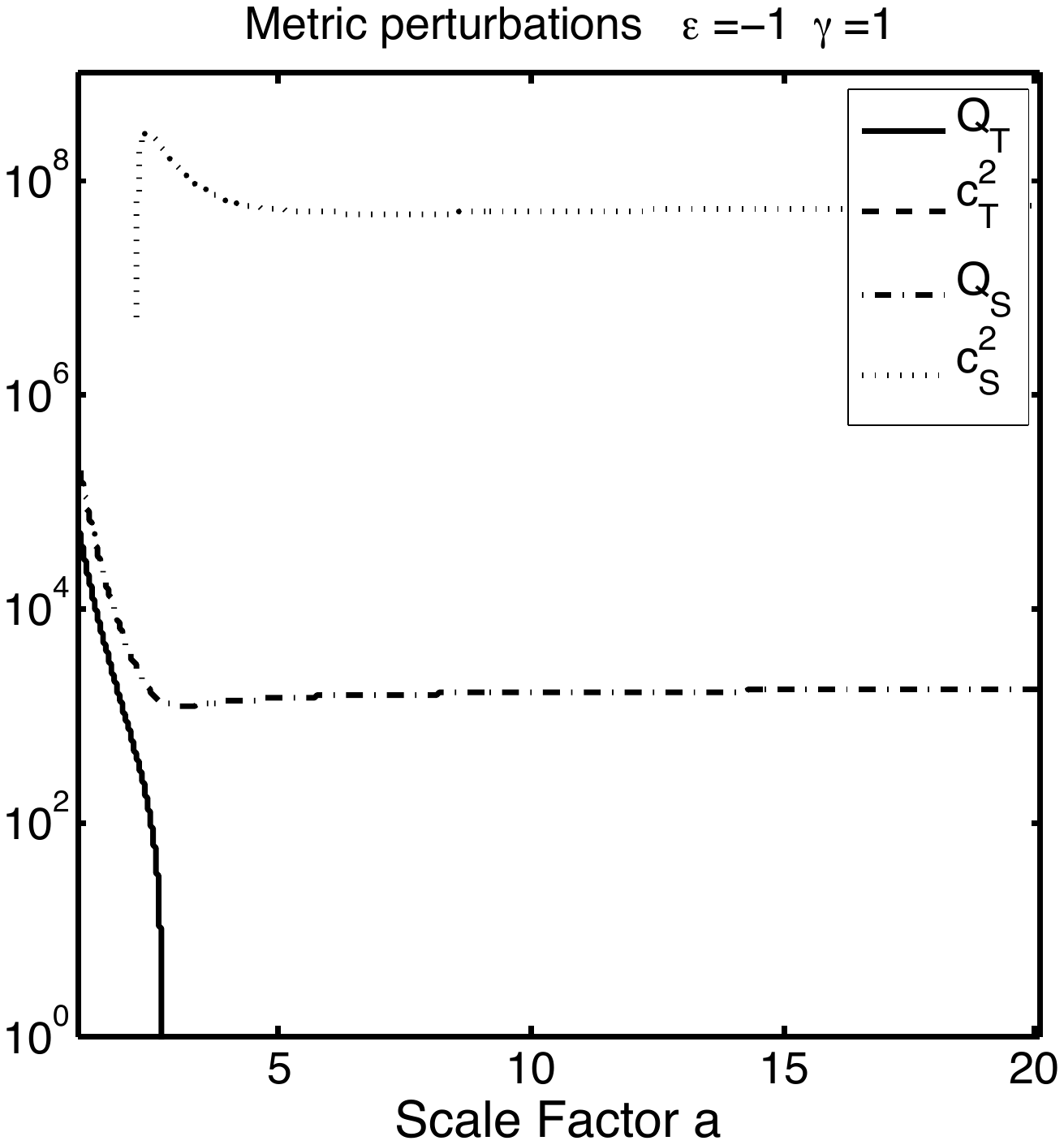}   \\
\includegraphics[width=0.35\textwidth]{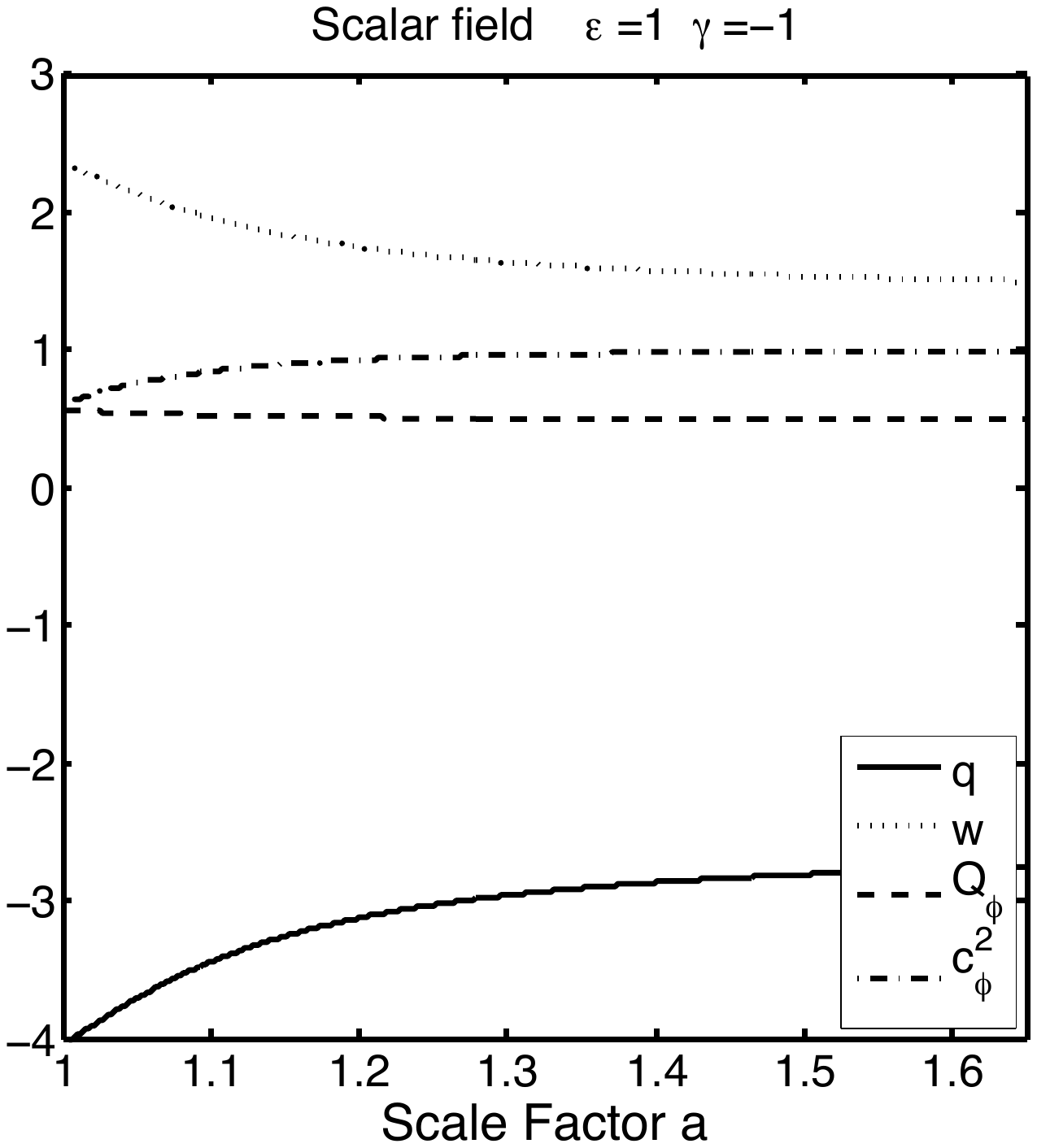}\hspace{1.5cm} 
\includegraphics[width=0.35\textwidth]{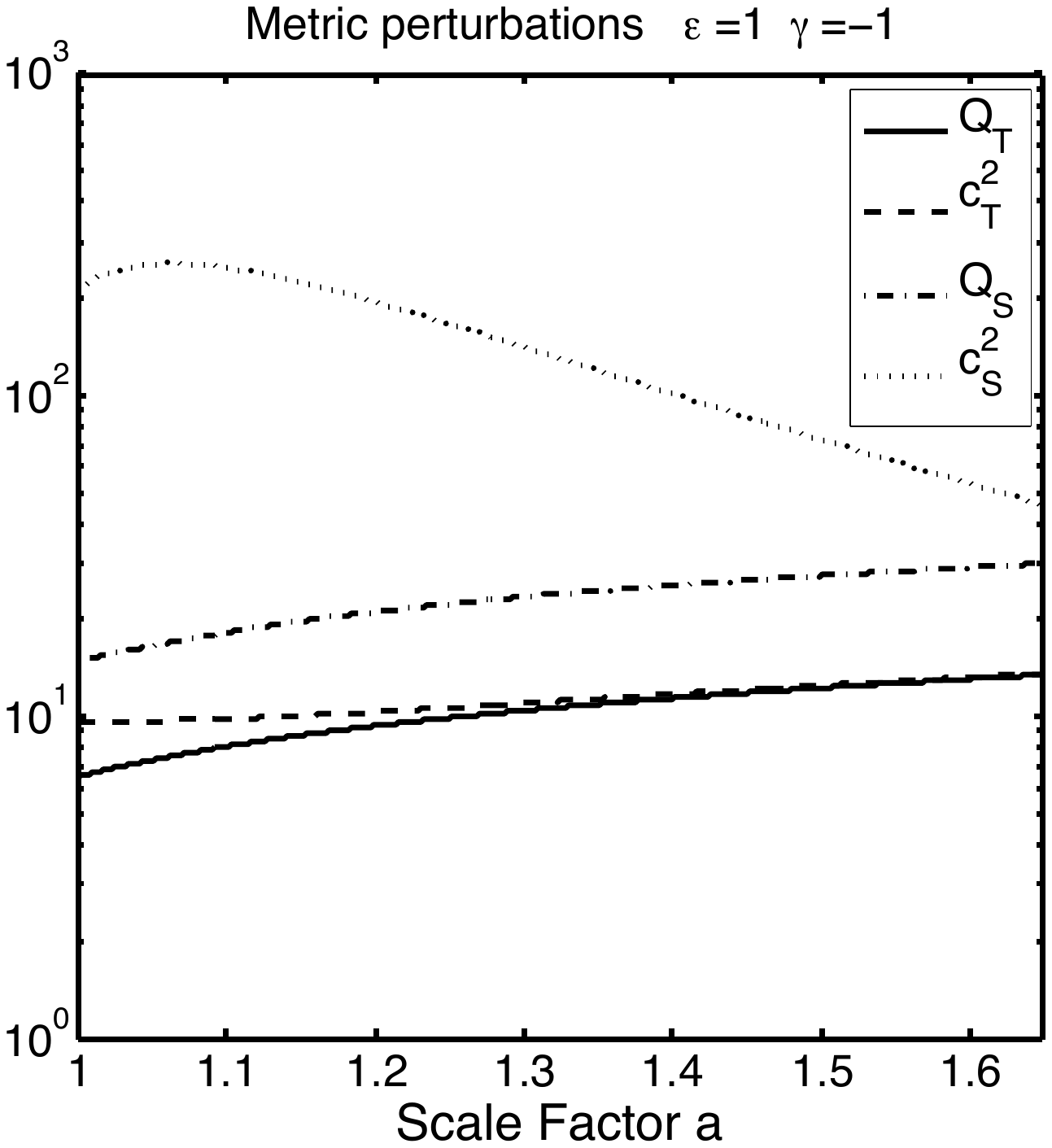} 
\end{center}      
\caption{ Cosmological evolution of $q, Q_{\phi}, c_{\phi}^2, Q_{S}, c_{S}^2, Q_{T}, c_{T}^2$ with the scale factor. (Top): the case $\epsilon=-1$ and $\gamma=1$ corresponds to a "double inflation". The scenario is pathological in many respects: singularity in the scalar field velocity (see top left) while the scalar field EoS is imaginary. Also (see top right, the $y$ axis is logarithmic): $c_T^2 <0$ and there are periods for which $c_s^2<0$ and $Q_T<0$. (Bottom): the case $\epsilon=1$ and $\gamma=-1$. The model is well behaved but does not accelerate the expansion, ie. as in the John alone model. The universe is actually in a super-stiff regime, and hence in a highly decelerating phase.}                    
\label{jgfigs}  
\end{figure*}

The case $\epsilon =-1$, $\gamma=1$ is clearly pathological for the various no-ghosts and no-acausal conditions, as seen on Fig.~(\ref{jgfigs}). Actually this theory leads to a double inflation scenario (see the acceleration parameter): the Universe transits from one de Sitter phase to another one, and experiences in between a super-acceleration phase. 
Finally, the case with negative $\gamma$ is similar to what we found for John alone: the theory is well-defined, ghost free and causal, but fail to exhibit any acceleration at all, see Figs.~(\ref{jgfigs}).
\subsection{Solar system}
\label{sec3c}
In this section we show how to derive solar system constraints on the free parameters of the model ``John + George''. 

In a first step we  solve asymptotically the field equations in spherical symmetric and isotropic coordinates. The line element reads 
\begin{equation} \label{sun:metric}
 ds^2= -A(r)^2 dt^2 + B(r)^2\left( dr^2 + r^2 d^2\Omega   \right),
\end{equation}
and the equations of motion for the components and for $\varphi$ are given in Appendix~\ref{app:sphericEOM}. The post-Newtonian analysis begins with the expansion of the metric components as $A^2=1 + \sum_i a_i/r^i$, $B^2=1+\sum_i b_i/r^i$ and $\varphi=p_0 + \sum_i p_i/r^i$ and of the field equations in powers of $1/r$. By equating the coefficients of equal powers of $r$ we find
\begin{subequations}              \label{sun:metcoeff}
	\begin{eqnarray}
		A^2 & = & 1-\frac{r_s}{r}+\frac{r_s^2}{2r^2} +\frac{\epsilon^2p_1^2}{4M_p^2z^2r^2}+\frac{p_1\epsilon r_s^2}{24M_pzr^3}\\
		&&-\frac{p_1^2 r_s}{24M_p^2z r^3} + \frac{3}{4}\frac{\bar\gamma\epsilon^2}{M_p^4z^2r^4}-\frac{r_s\bar \gamma }{8M_p^4zr^5},\non \\
		B^2&=&1+\frac{r_s}{r} - 2\frac{\epsilon p_1}{M_pzr} -\frac{p_1^2 }{4M_p^2zr^2}-\frac{\bar \gamma}{4M_p^4zr^4},
	\end{eqnarray}
\end{subequations}
where $r_s=\frac{2GM}{c^2}$ is the Schwarzschild radius of the central body, $z=1+\frac{\epsilon p_0}{M_p}$, and $\bar \gamma=\gamma p_1^2$. In the expansion above, we neglected  higher order terms in $r_s/r$, in $\epsilon$, in $p_1/(rM_p)$ and in $\bg/(rM_p)$ (which means we suppose these terms to be smaller than 1). We recall that, in our conventions,  $\gamma$ and $\epsilon$ are dimensionless parameters. The asymptotic scalar field value  $p_0=\phi(r\to \infty)$ (in GeV) is a free parameter that can eventually be connected to the cosmological evolution of the scalar field. 

The next step then amounts to relate the dimensionless parameter $p_1$ to the scalar charge of the central body, and thus eventually to the couplings $\epsilon$ and $\gamma$. This could be determined numerically by solving the equations in the interior of the body \cite{Damour:1993hw} with suitable asymptotic and regularity conditions, together with, e. g. a polytropic equation of state for the star's interior. Such a discussion about possible spontaneous scalarization goes however beyond the scope of the present paper and is left for future studies. In this section we only sketch how solar systeme constraints can be derived.

The final step consists then in computing observables effects from the metric Eqs.~(\ref{sun:metric}-\ref{sun:metcoeff}), which show a deviation from standard General Relativity in the Solar System. A first constraint comes from the anomalous perihelion shift. First of all, we determine the geodesic equation (to first order in the metric deviation) to find the planetary equations of motion\footnote{When computing the geodesics, we suppose that matter is minimally coupled to the metric.}.  The anomalous term  is treated perturbatively to find the perihelion shift with the Gauss equations, which determines the evolution of the different orbital parameters due to  perturbative forces. In particular, we derive the rate of change of  the argument of the perihelion $\omega$. Finally, the secular term of the change of the argument of the perihelion is exhibited by a time average (over an orbital period $T$)
\begin{equation}
 \left<\frac{d\omega}{dt}\right> = \frac{1}{T}\int_0^T     \frac{d\omega}{dt} dt.
\end{equation} 
The results of this whole procedure is given by
\begin{eqnarray} \label{domega}
\left<\frac{d\omega}{dt}\right> &=&-\alpha_1 \frac{\bg}{M_p^4z}-\alpha_2\frac{\bg \epsilon^2}{M_p^4z^2}  - \alpha_3 \frac{p_1^2}{M_p^2 z} \\
&&- \alpha_4 \frac{p_1\epsilon}{zM_p}-\alpha_5\frac{p_1^2\epsilon^2}{z^2M_p^2},\non
\end{eqnarray} 
with 
\begin{subequations}    \label{alphai}
\begin{eqnarray}
	\alpha_1 &=  &\frac{n}{16a^4}\frac{8+24e^2+3e^4}{(1-e^2)^4} , \\
	\alpha_2 &=&\frac{9n}{8a^3r_s}\frac{4-3e^2-e^4}{(1-e^2)^4},\\
	\alpha_3 &=& \frac{n}{8a^2}\frac{4+e^2}{(1-e^2)^2},\\
	\alpha_4&=& \frac{2n}{a(1-e^2)},\\
	\alpha_5&=& \frac{n}{4ar_s(1-e^2)}, 
\end{eqnarray}                         
\end{subequations}
where  $a$  is the  semi-major axis of the orbit, $e$ its eccentricity, and $n$ its mean motion $n=\frac{2\pi}{T}$. 
The advance of perihelion of Solar System planets is very tightly constrained by planetary ephemerides. In particular, INPOP10a gives constraints on supplementary advances of perihelia (see Table 5 of~\cite{fienga:2011qf}). Table~\ref{tableperihelie} gives the value of the $\alpha_i$ coefficients appearing in the expression of the advance of perihelion (\ref{domega}) and the constraints coming from INPOP10a.   

	    
\begin{table*}[htb]
\centering
	\begin{tabular}{|c|c|c|c|c|c|c|}\hline
&$\alpha_1$ ($mas/cy/m^4$) &  $\alpha_2$ ($mas/cy/m^4$) &$\alpha_3$ ($mas/cy/m^2$) &  $\alpha_4$ ($mas/cy/m$) &$\alpha_5$ ($mas/cy/m^2$) & $d\omega/dt$ ($mas/cy$) INPOP10a \\\hline 
Mercury  &  $3.2  \times 10^{-32}$  &  $4.85 \times 10^{-24}$   &  $8.83 \times 10^{-11}$  &   $19.4$   &  $8.2 \times 10^{-4}$   &     $0.4\pm0.6$ \\
Venus    &  $7.69 \times 10^{-34}$  &  $2.53 \times 10^{-25}$   &  $9 \times 10^{-12}$     &   $3.89$            &  $1.65 \times 10^{-4}$  &     $0.2\pm1.5$ \\ 
Earth    &  $1.3  \times 10^{-34}$  &  $5.9  \times 10^{-26}$   &  $2.9 \times 10^{-12}$   &   $1.73$            &  $7.33 \times 10^{-5}$  &     $-0.2\pm0.9$ \\ 
Mars     &  $1.36 \times 10^{-35}$  &  $9.13 \times 10^{-27}$   &  $6.77 \times 10^{-13}$  &   $6.1 \times 10^{-1}$   &  $2.58 \times 10^{-5}$  &     $-0.04\pm 0.15$ \\ 
Jupiter  &  $1.51 \times 10^{-38}$  &  $3.56 \times 10^{-29}$   &  $9.06 \times 10^{-15}$  &   $2.81 \times 10^{-2}$  &  $1.19 \times 10^{-6}$  &     $-41\pm42$ \\ 
Saturn   &  $5.33 \times 10^{-40}$  &  $2.3  \times 10^{-30}$   &  $1.08 \times 10^{-15}$  &   $6.16 \times 10^{-3}$  &  $2.61 \times 10^{-7}$  &     $0.15\pm0.65$ \\ 
\hline	\end{tabular}
\caption{Values of the different $\alpha_i$ coefficients (\ref{alphai}) for planets and constraints given by INPOP10a for supplementary advances of perihelia (values given in~\cite{fienga:2011qf}).}
\label{tableperihelie}
\end{table*}          

The most stringent constraints in the case considered here are obtained by  data from Mercury and read
\begin{subequations}   \label{const_per}
\begin{eqnarray}
-3.12 \times 10^{31} \ m^{4}   & < \frac{\bg}{M_p^4z} <     & 6.25 \times 10^{30} \  m^{4}\vspace*{5mm}\label{constgamma}, \\
-2.06 \times 10^{23} \ m^{4}   & < \frac{\bg \epsilon^2}{M_p^4z^2} <  & 4.12 \times 10^{22} \ m^{4}\vspace*{5mm}, \\ 
-1.13 \times 10^{10} \ m^{2}   & < \frac{p_1^2}{M_p^2 z} <                                & 2.26 \times 10^{9}  \ m^{2}\vspace*{5mm}, \\ 
-5.16 \times 10^{-2} \  m       & < \frac{p_1\epsilon}{zM_p} <                        & 1.03 \times 10^{-2} \ m.
\end{eqnarray}     
\end{subequations}

\begin{figure}[t]
\begin{center}
\includegraphics[width=9cm]{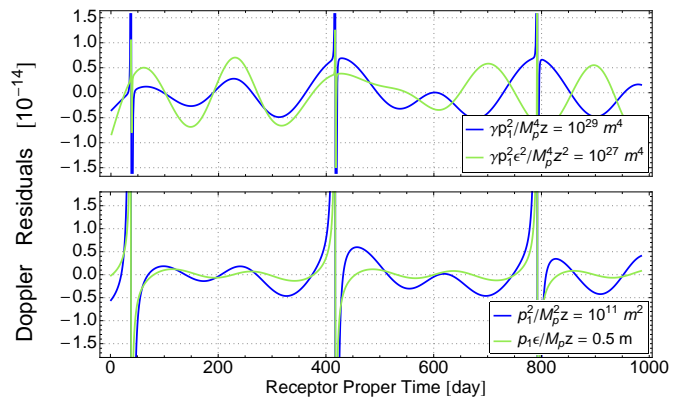}
\end{center}
\caption{Incompressible anomalous signal due to the different terms in the metric (\ref{sun:metric}-\ref{sun:metcoeff}) on Cassini Doppler signal with indicated values. The produced residuals are larger than Cassini accuracy (which is $10^{-14}$), therefore these values for the parameters are ruled out.}
\label{figdoppler}
\end{figure}

These constraints are obtained by considering only deviations  on the dynamics (i.e. on the equations of motion). Other constraints can be derived using propagation of light rays in the solar system. For example, radioscience experiments include light propagation through a Shapiro-like term that can be derived from the expression of the metric (\ref{sun:metric}-\ref{sun:metcoeff}). One way to obtain such constraint is to follow the strategy and to use the software of \cite{hees:2012fk,hees:2011vn,hees:2011kx}. The main idea presented in these papers is to simulate radioscience observables directly from the space-time metric so that the software includes both deviations on the dynamics and deviations on light propagation. In particular, we use this software to simulate a two-way Doppler link between Earth and Cassini spacecraft from May 2001 on the metric  (\ref{sun:metric}-\ref{sun:metcoeff}) and to analyze them in GR by fitting the initial conditions of the spacecraft. The residuals that emerge are the incompressible signature produced by the alternative theory considered on Doppler signal for Cassini. Comparing this signal to the Doppler accuracy of the mission ($10^{-14}$) allows us to give order of magnitude of constraints on the theory. Fig.~\ref{figdoppler} represents the incompressible signatures produced by parameters entering the metric (\ref{sun:metric}-\ref{sun:metcoeff}) on Cassini Doppler. The three sharp peaks occurred at solar conjunctions. The order of magnitude of the residuals observed in this figure is larger than Cassini accuracy, which means that the values of the parameters should be smaller than the indicated values.

\begin{figure}[htb]
\begin{center}
 \includegraphics[width=0.235\textwidth]{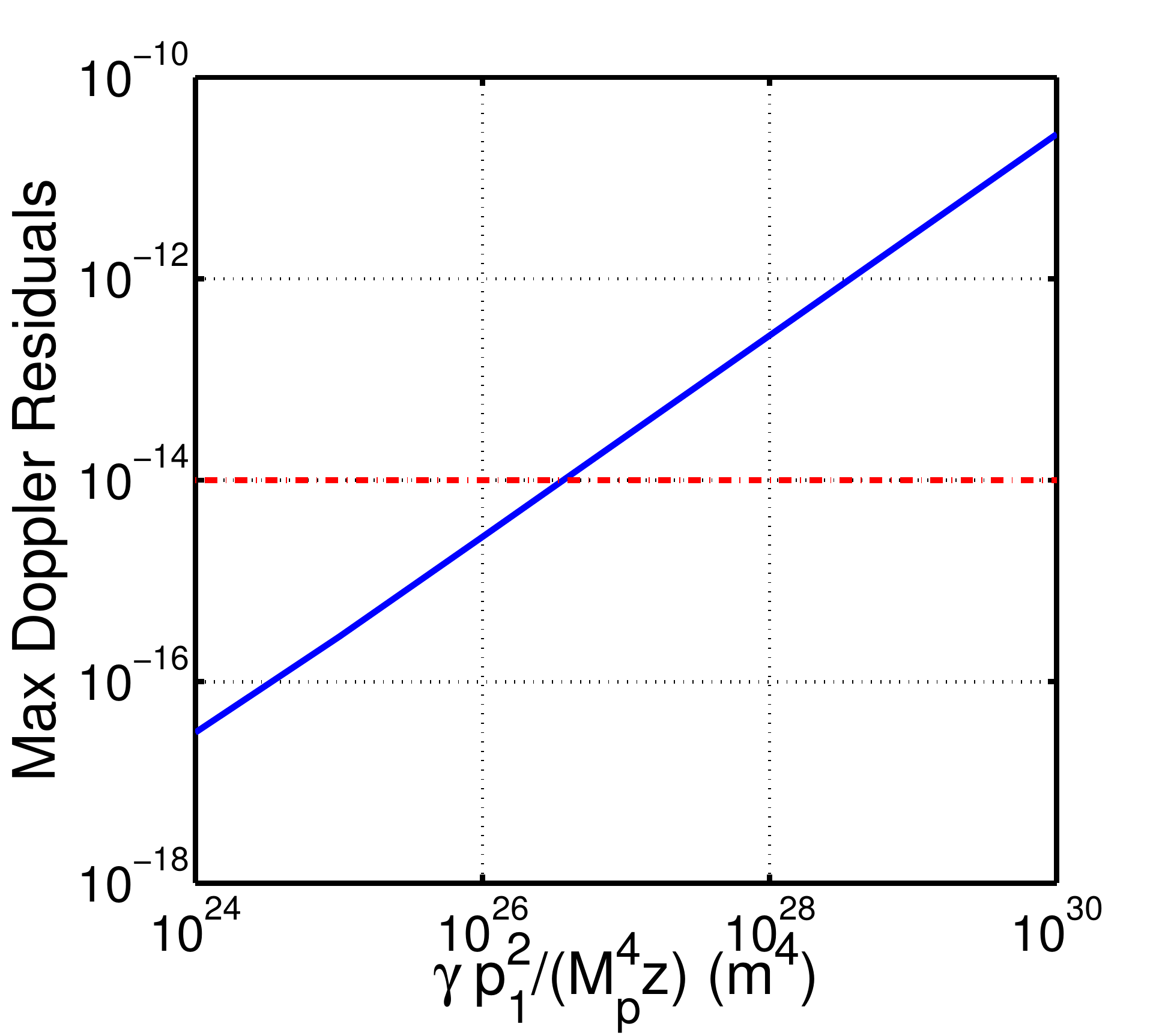} \hfill
 \includegraphics[width=0.235\textwidth]{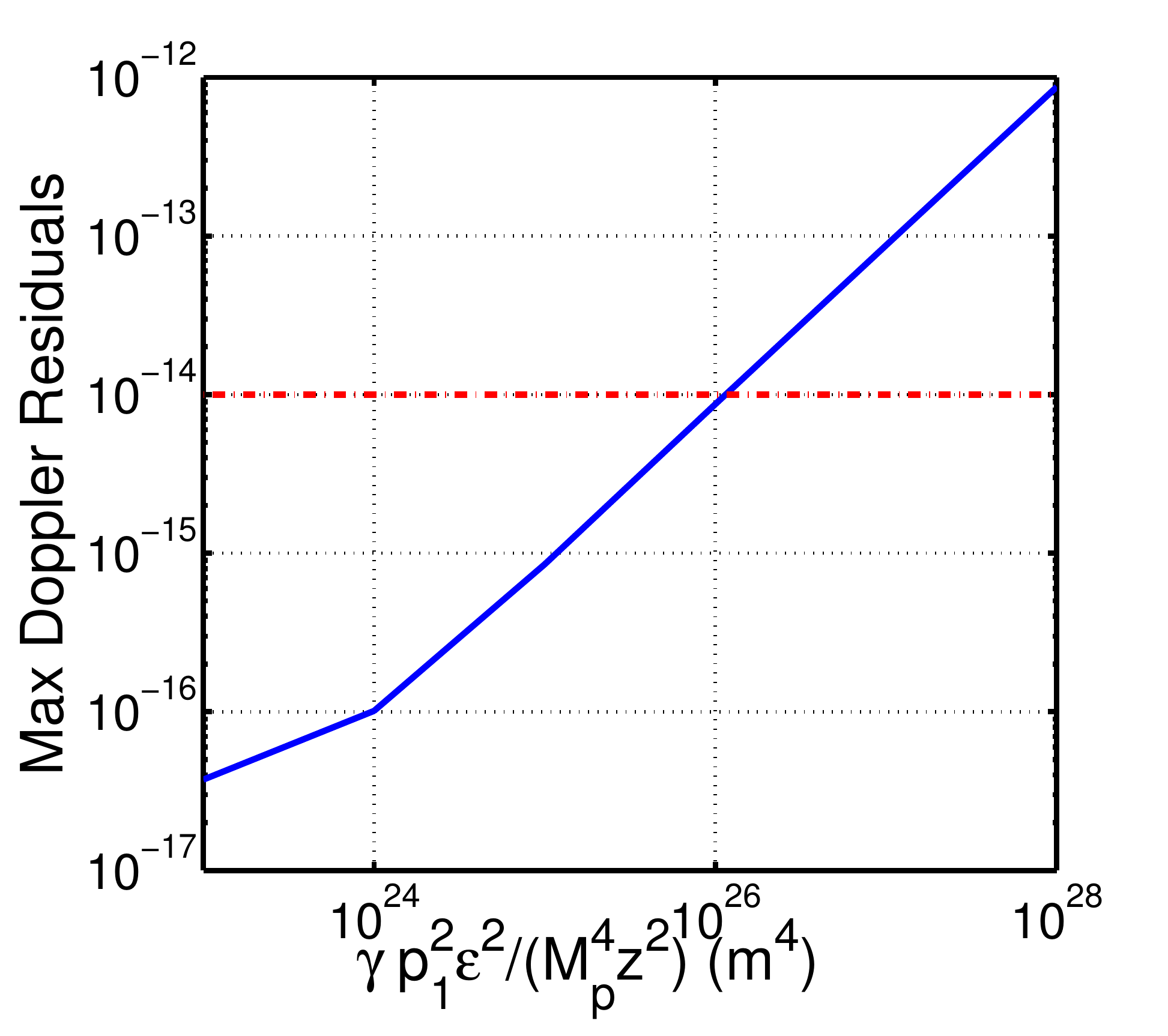}
 \includegraphics[width=0.235\textwidth]{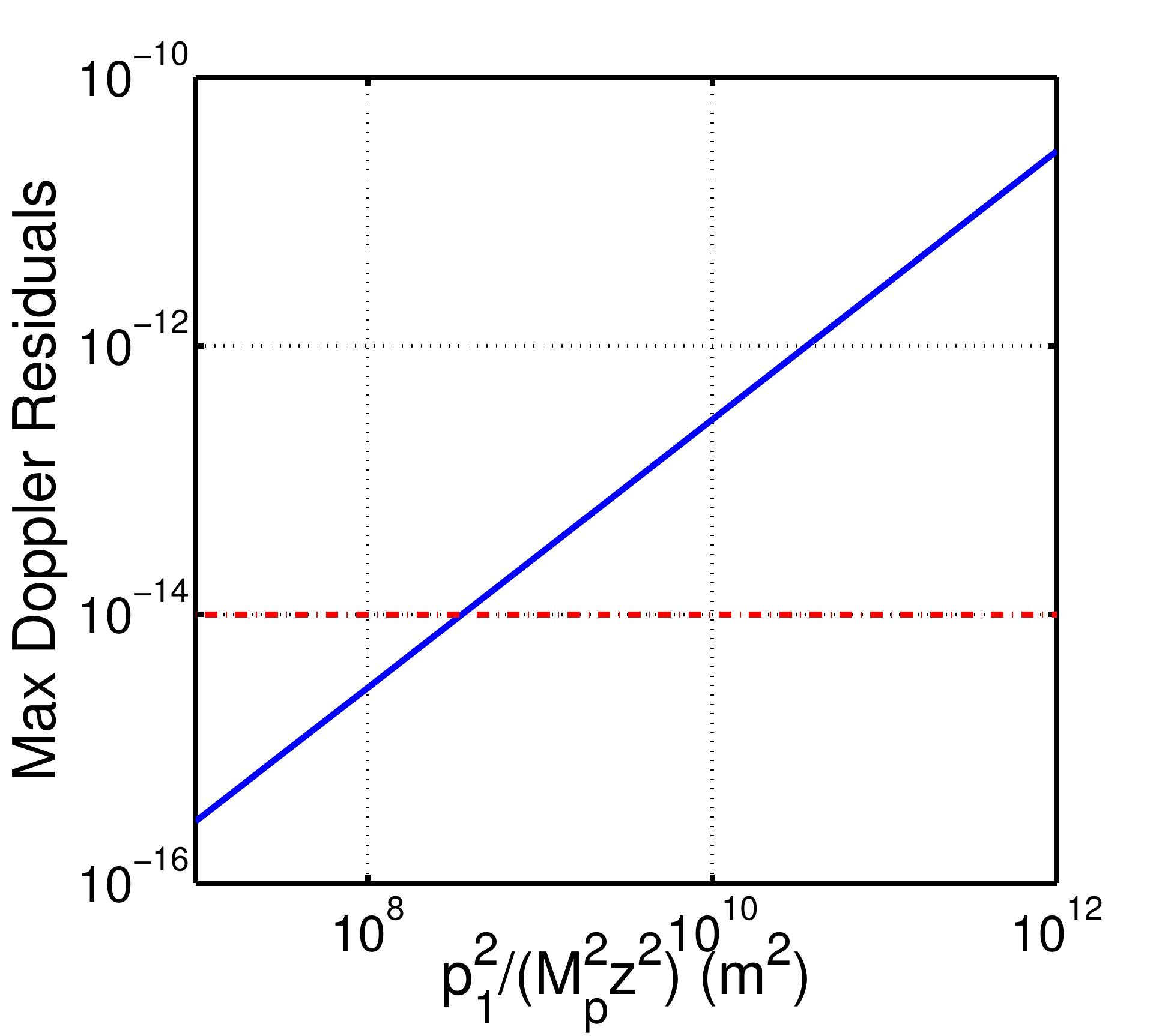}\hfill
 \includegraphics[width=0.235\textwidth]{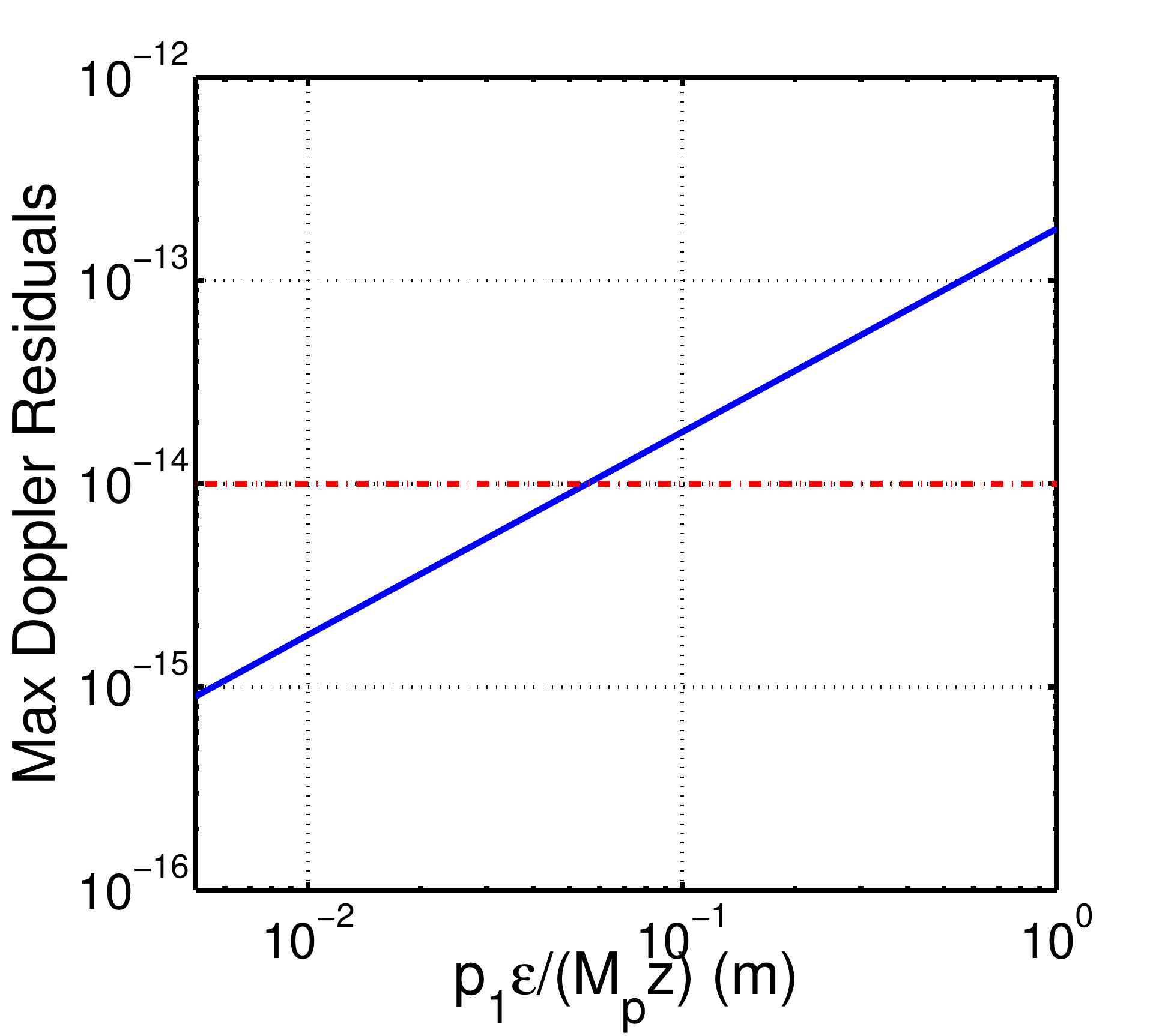}
\end{center}
\caption{Representation of the maximal  Doppler residuals signal due to John+George modification of gravity parametrized by the four parameters  in the metric (\ref{sun:metric}-\ref{sun:metcoeff})  for the Cassini mission between Jupiter
  and Saturn. The red (dashed) lines represent the assumed Cassini accuracy.}
\label{figCass}
\end{figure}

We run a set of simulations with different values for the parameters appearing in the metric (\ref{sun:metric}-\ref{sun:metcoeff}). Fig.~\ref{figCass} represents the evolution of the maximal Doppler residuals obtained in Cassini signal as function of metric parameters. Requiring the residuals to be lower than Cassini accuracy ($10^{-14}$) gives the boundary values
\begin{subequations}
\begin{eqnarray}
  \left| \frac{\bg}{M_p^4z} \right|= \left|\frac{\gamma p_1^2}{M_p^4(1+\frac{\epsilon p_0}{M_p})} \right|   & <  & 3.65 \times 10^{26} \ m^4, \\ 
\left| \frac{\bg \epsilon^2}{M_p^4z^2} \right|= \left|\frac{\gamma p_1^2 \epsilon^2}{M_p^4(1+\frac{\epsilon p_0}{M_p})^2} \right| & < &  1.15 \times 10^{26} \ m^4, \\
\left| \frac{p_1^2}{M_p^2 z} \right|= \left|\frac{p_1^2}{M_p^2(1+\frac{\epsilon p_0}{M_p})} \right|   &<& 3.53 \times 10^{8} \ m^2, \\
\left|\frac{p_1\epsilon}{zM_p} \right|= \left|\frac{p_1\epsilon}{M_p(1+\frac{\epsilon p_0}{M_p})} \right| &<& 5.56 \times 10^{-2} \ m    .
\end{eqnarray}	
\end{subequations}
It should be noted that radioscience constraints are significantly better for $\frac{\bg}{M_p^4z}$ and $\frac{p_1^2}{M_p^2 z}$ as compared to perihelia advances. On the other hand, the constraint from the ephemerides on $\frac{\bg \epsilon^2}{M_p^4z^2}$ is significantly better while the constraint on $\frac{p_1\epsilon}{zM_p}$ is of the same order of magnitude. 

We thus see that solar systems tests of GR will put severe constraints on the free parameters of the model. As already stressed however, the full numerical study of Solar's interior is required, while a thorough analysis of possible spontaneous scalarization (depending also of the Georges coupling) is necessary.
\section{Conclusions}

\noindent In this paper, we have explored some phenomenology associated to a subset of the ``Fab Four'' scalar-tensor theory. The philosophy behind this preliminary work is that we cannot forget about solar system constraints on the parameter space, even when we deal with inflationary solutions. Traditional models of inflation rely upon the fact that the inflaton field decays, at some stage, into ordinary matter through some reheating mechanism. Therefore, the scalar-tensor nature of inflationary gravity is lost very soon in the evolution of the Universe. On the opposite, in the models studied in this paper the scalar field should live and show its effects until nowadays. Therefore, the parameter space determined by constraints from cosmological observations must overlap with the one determined by solar system tests. The ``Fab Four'' theory has many parameters with a very rich phenomenology, and with this paper we begin an ambitious plan for its systematic study.

In this work, we contained ourselves to the cases John and John plus George. The John case represents a theory with a non-minimal derivative coupling between scalar field and Einstein tensor. It was already known that this model admits inflationary solutions with a graceful exit. Here, we constrained the coupling constant $\gamma$ by showing that it must be positive in order to have successful inflation. If this is the case, however, very unnatural initial conditions are required. In particular the field velocity, which is related to the energy density, must be huge compared to the Planck scale, rendering the theory no longer trustworthy.  Negative values for $\gamma$ are permitted but do not allow for inflation while ghost states might appear. The most serious problem however comes from the fact that the model  turns out to be trivial when one tries to solve the equations of motion inside a compact object. Indeed, we found that the only solution with a finite field at the centre is  $\varphi=0$ everywhere.

These facts have convinced us to extend the theory to include the term named ``George'', which is nothing but a coupling between the scalar field and the Ricci scalar. The parameter space is now two-dimensional and we solved numerically the equations of motion for a cosmological background. The main result is that the sign of the two coupling constants must be positive in order to have an inflationary phase with graceful exit. We reserve the analysis of the naturalness problem, i.e. the need for extreme initial conditions, for future work. 
We also performed a post-Newtonian analysis of the theory by solving the equations of motion by imposing static and spherical symmetry and expanding the fields for large radial distance. The aim was to put some constraint on the free parameters, which now include also the asymptotic value of the scalar field and the scalar charge, which now makes sense as there are non-trivial solution also for the interior of a compact object. The results still allow for a large parameter space, therefore future work is necessary in order to improve the constraints.

As mentioned above, this paper is the first of a series that aim at a systematic study of the ``Fab Four'' phenomenology, in order to isolated the experimentally viable sectors of the theory. Besides these aspects, there are several issues that deserve further investigations. For example, if these models truly lead to inflationary cosmology and graceful exit, we will need an alternative reheating mechanism. Maybe, the complexity of the theory acts as an effective potential for the scalar field, which resembles the usual power-law forms of usual inflation. But if this is not the case, then we need to find alternative explanations.  Another aspect that needs to be studied is the relevance of these modifications of gravity in terms of late-time cosmology, as we expect modifications driven by the scalar field. This would include, at the background level, the study of tracking solutions and of the convergence mechanism towards GR, if any. This has not yet been adressed in the litterature for the John Lagrangian (but see \cite{Copeland:2012qf}). The study of cosmological perturbations, in particular CMB spectra and large scale structures, might then further reduce the parameter space. Finally, gravitational effects might be relevant at galactic scales and give rise to alternative explanations to the anomalous galactic rotation curves.

In conclusion, we believe that the recent developments in scalar-tensor theories of gravity have opened the door to new and intriguing research directions, and we are confident that many interesting results will be obtained in the near future.

\begin{acknowledgments}
J.-P. B. is FSR/COFUND postdoctoral researcher at naXys.
M. R. is supported by a grant of ARC 11/15-040 convention.
A. K. thanks ARC 11/15-040 for travel support.
A. Hees is FNRS Research Fellow. A. Hees thanks B. Lamine, P. Wolf, C. Le Poncin-Lafitte, V. Lainey, S. Reynaud and M.T. Jaekel for useful discussions about radioscience simulations within Solar System.
\end{acknowledgments}

\appendix
\begin{widetext}
\section{Cosmological equations}
\label{AppA}
\noindent  In terms of the reduced variables $x(t)=\kappa \dot{\phi}$, $y(t)=\sqrt{\kappa} \dot{\alpha}$ and $z(t)= 1 + \epsilon \sqrt{\kappa} \phi(t)$, the equations of motion for a flat and empty universe derived from action Eq.~(\ref{actionjohngeorges}) are 
\begin{eqnarray}
&&6 \epsilon x y+x^2 \left(-1+9 \gamma y^2\right)+6 y^2 z=0 \label{Eomjg1} \\
&&4 x \left(\epsilon+\gamma \sqrt{\kappa} \dot{x}\right) y+x^2 \left(1+3 \gamma y^2+2 \gamma \sqrt{\kappa} \dot{y}\right)+6 y^2 z+2 \sqrt{\kappa} (\epsilon \dot{x}+2 \dot{y} z)=0  \label{Eomjg2} \\
&& 3 y \left(x-2 \epsilon y-3 \gamma x y^2\right)+\sqrt{\kappa} \left(\dot{x}-3 \gamma \dot{x} y^2-3 (\epsilon+2 \gamma x y) \dot{y}\right)=0 \label{Eomjg3}
\end{eqnarray}
which can be decoupled in the following way:
\begin{eqnarray}
&&\dot{x}=\frac{-3 x \left[\epsilon x+4 \left(\epsilon^2+\gamma x^2\right) y+7 \epsilon \gamma x y^2\right]+6 y (-2 x+\epsilon y) z}{2 \kappa^{1/2} \left(3 \epsilon^2+12 \epsilon \gamma x y+\gamma x^2 \left(1+9 \gamma y^2\right)+\left(2-6 \gamma y^2\right) z\right)}\\
&&\dot{y}=\frac{2 \epsilon x y \left(1-15 \gamma y^2\right)+x^2 \left[-1+3 \gamma y^2 \left(4-9 \gamma y^2\right)\right]-6 y^2 \left(2 \epsilon^2+z-3 \gamma y^2 z\right)}{2 \kappa^{1/2} \left[3 \epsilon^2+12 \epsilon \gamma x y+\gamma x^2 \left(1+9 \gamma y^2\right)+\left(2-6 \gamma y^2\right) z\right]}
\end{eqnarray}
The scalar field EoS is given by:
\beq
w_{\phi}= - x \left(3 \gamma x^2+2 z\right) \frac{N}{D}
\eeq
with
\begin{eqnarray}
N&=&6 x \left(\gamma x^2-z\right) \left(3 \gamma x^2+2 z\right)^2+6 \sqrt{3} \epsilon^3 \sqrt{x^2 \left(3 \epsilon^2+3 \gamma x^2+2 z\right)} \left(7 \gamma x^2+2 z\right)-18 \epsilon^4 \left(7 \gamma x^3+2 x z\right)\nonumber\\&&-2 \sqrt{3} \epsilon \sqrt{x^2 \left(3 \epsilon^2+3 \gamma x^2+2 z\right)} \left(33 \gamma^2 x^4+16 \gamma x^2 z-4 z^2\right)+9 \epsilon^2 \left(15 \gamma^2 x^5+4 \gamma x^3 z-4 x z^2\right)\\
D&=&\left[-3 \epsilon x+\sqrt{3} \sqrt{x^2 \left(3 \epsilon^2+3 \gamma x^2+2 z\right)}\right]^2 \times \nonumber\\ &&\left[18 \gamma^3 x^6+30 \gamma^2 x^4 z+24 \gamma x^2 z^2+8 z^3+6 \sqrt{3} \epsilon \gamma x \left(\gamma x^2+2 z\right) \sqrt{x^2 \left(3 \epsilon^2+3 \gamma x^2+2 z\right)}+3 \epsilon^2 \left(3 \gamma^2 x^4+4 z^2\right)\right] \nonumber\\
\end{eqnarray}

\section{Ghosts conditions}
\label{AppB}
\noindent The coupling to the Ricci scalar, "George", does not change the analysis made for the scalar field sector of the theory. Thus the two following conditions still hold 
\begin{eqnarray}
&&Q_{\phi}=\frac{1}{2}\left(1-3 \gamma y^2 \right) >0 \\
&&c_{\phi}^2= \frac{1 - \gamma \left(3 y^2 +2 \sqrt{\kappa} \dot{y} \right)}{1-3  \gamma y^2 } \geq 0
\end{eqnarray}
For the metric perturbations, we derive, based on  Eqs.~(23), (25), (26), and (27) of \cite{DeFelice:2011bh}
\begin{eqnarray}
&&Q_T>0 \Rightarrow z + \frac{\gamma x^2}{2} >0 \\
&&c_T^2 \geq 0 \Rightarrow z - \frac{\gamma x^2}{2} \geq 0\\
\end{eqnarray}
for the tensorial part, and also 
\beq
Q_S>0 \Rightarrow 3 \epsilon^2+12 \epsilon \gamma x y+9 \gamma^2 x^2 y^2+2 z+\gamma \left(x^2-6 y^2 z\right)>0
\eeq
for the scalar part of the metric perturbations, while their squared speed $c_s^2 \geq 0$ leads to \begin{eqnarray}
&&2 y \left(\frac{\gamma x^2}{2}+z\right)^2 \left(\epsilon x+3 \gamma x^2 y+2 y z\right)+2 x \left(\epsilon+\gamma \sqrt{\kappa} \dot{x}\right) \left(\gamma x^2+2 z\right) \left(\epsilon x+3 \gamma x^2 y+2 y z\right) \nonumber \\
&&+\frac{1}{2} \left(\gamma x^2-2 z\right) \left(\epsilon x+3 \gamma x^2 y+2 y z\right)^2 \nonumber \\
&&-2 \sqrt{\kappa} \left(\frac{\gamma x^2}{2}+z\right)^2 \left[\epsilon \left(\dot{x}+\frac{2 x y}{\sqrt{\kappa}}\right)+3 \gamma x (2 \dot{x} y+x \dot{y})+2 \dot{y} z\right] \geq 0
\end{eqnarray}

\section{Spherically symmetric equations of motion}\label{app:sphericEOM}

\noindent We derive the equations of motion for the action (\ref{actionjohngeorges}) with a spherically symmetric and static field configuration. We consider the metric (\ref{sun:metric}), and  we replace its components in the Lagrangian. With the Noether theorem,  we find the equations of motion for the fields $A$, $B$, $R$, and $\varphi$. Finally, we impose the gauge  $R=r$, and we find three equations plus a Hamiltonian constraint that read 
\begin{eqnarray}
0&=&\Big( 2\varphi'^{2}B{r}^{2}B'' + 4\varphi'  B^2r\varphi'' +4 \varphi' B{r}^{2}  \varphi''  B' -5 \varphi'^{2}B'^{2}{r}^{2}+2\varphi'^2B^2 \Big) \gamma\kappa^{2}+\\\non
&+& \Big( -4B^{3}\varphi   B''  {r}^{2}-2 B^{3} B'{r}^{2}\varphi' -4B^{4} \varphi' r-8B^{3}\varphi   B' r-2B^{4}{r}^{2}\varphi''  +2 B^2\varphi   B'^{2}{r}^{2} \Big) {\epsilon \kappa^{1/2}}+\\\non
&-&8B^{3}rB' -4B^{3}{r}^{2}B'' +2 B^2{r}^{2} B'^{2}- \varphi'^{2}{\kappa}^{2}B^{4}{r}^{2},\\\non\\
0&=&\Big( 4 \varphi'^{2} B' A rB +2\varphi'^2B^2A  -3 \varphi'^{2} B'^{2}A  {r}^{2}+4 \varphi' B {r}^{2}\varphi''A B'  +8\varphi'^2B^2rA'  +2\varphi'^2B'  A' B {r}^{2}+\\\non
&+&2 \varphi'^{2} A'' {r}^{2} B^2+4 \varphi' B^{2}r \varphi'' A  +4 \varphi'  A' {r}^{2} \varphi''  B^2+2 \varphi'^{2}A B  {r}^{2}B''  \Big) \gamma\kappa^{2}+ \\\non
&+&\Big( -4B^{3}A   B' {r}^{2}\varphi' +2 B^{2}A \varphi   B'^{2}{r}^{2}-8B^{4}A  \varphi' r-8\varphi  r A' B^{4}-8B^{3}A \varphi   B'r-4 B^{3}A \varphi   B'' {r}^{2}\\\non
&-&4\varphi   B'  A' {r}^{2}B^{3}-6 A'{r}^{2} \varphi' B^{4}-4\varphi A'' {r}^{2}B^{4}-4B^{4}A  {r}^{2}\varphi''   \Big) \epsilon\kappa^{1/2}+\\\non
&-&4B^{3}A  {r}^{2}B'' -4{r}^{2} A'' B^{4}-8r A'B^{4}-8B^{3}A  rB' -4 A' {r}^{2} B'B^{3}+2 B^2A  {r}^{2}  B'^{2}- \varphi'^{2}{\kappa}^{2}B^{4}A {r}^{2},\\\non\\
0&=&\Big( 4 \varphi'  A'  B^{3}+4 \varphi' A  B'  B^2+4 \varphi'  B' A {r}^{2} B''B +4 \varphi'  B''A' {r}^{2}B^{2}+4 \varphi'' r A'  B^{3}+4 \varphi' r A'' B^{3}+\\\non
&+&4 \varphi'' B'A' {r}^{2}B^{2}+8 \varphi'  B'A' r B^2+4 \varphi'  B''A  r B^2+4\varphi'' B'A r B^{2}\\\non
&-&4 \varphi'  B'^{2}A rB -6 \varphi'  B'^{2} A'{r}^{2}B -6 \varphi'  B'^{3}A {r}^{2}+4 \varphi'  B' A'' {r}^{2} B^2+2\varphi'' B'^{2}A {r}^{2}B   \Big) \gamma\kappa+ \\\non
&+&\Big( -8 B' A r B^{4}-4B^{5}rA' -4{r}^{2}A   B''
B^{4}-2 B'A' {r}^{2}B^{4}-2B^{5}{r}^{2}A'' +2 B'^{2}A {r}^{2} B^{3} \Big) \epsilon\kappa^{-1/2}\\\non
&+&2 A'{r}^{2} \varphi' B^{5}+2B^{5}{r}^{2}\varphi''A+4 \varphi' B^{5}A r+2 B' {r}^{2} \varphi'A B^{4},\\\non\\
0&=&\Big( 2 \varphi'^{2} A'  B^{2}+2r \varphi'^{2} A''  B^2-6 \varphi'^{2} B'^{2}A r+4 \varphi'A  \varphi''  B^2-2 \varphi'^{2} B' A B-4\varphi'^2A'B'rB\\\non
& +& 4 \varphi' r B'A  \varphi'' B +4r \varphi' A'  \varphi''  B^2+2 \varphi'^{2}r B''A B \Big) \gamma\kappa^{2}+ \Big( -4\varphi   B'A B^{3}+4r\varphi   B'^{2}A   B^2+\\\non
&-&4 B^{4}rA' \varphi' -4 B^{4}\varphi  A'' r-4 B^{4}\varphi''A r-4 B^{4}\varphi  A' -4B^{4}\varphi' A  -4\varphi   B''A r B^{3} \Big) \epsilon\kappa^{1/2}\\\non
&+&4r B'^{2}A  B^{2}-4rA  B'' B^{3}-4 B' A B^{3}-4B^{4}A' -4B^{4} A'' r-2 B^{4} \varphi'^{2}A r{\kappa}^{2}.
\end{eqnarray}

\end{widetext}

\bibliography{mybib}

\end{document}